\documentclass{aa}  

\usepackage{graphicx}
\usepackage{amsmath}
\usepackage{array}
\usepackage{soul}
\usepackage{txfonts}
\usepackage{booktabs}
\usepackage{hyperref}
\usepackage{tabularx}
\hypersetup{
	colorlinks=true,
	linkcolor=blue,
	citecolor=blue,
	urlcolor=cyan
}

\newcommand{\Ca}[1]{\ensuremath{^{#1}\mathrm{Ca}}}
\newcommand{\Ti}[1]{\ensuremath{^{#1}\mathrm{Ti}}}
\newcommand{\Cr}[1]{\ensuremath{^{#1}\mathrm{Cr}}}
\newcommand{\Mn}[1]{\ensuremath{^{#1}\mathrm{Mn}}}
\newcommand{\Fe}[1]{\ensuremath{^{#1}\mathrm{Fe}}}

\newcommand{\Zn}[1]{\ensuremath{^{#1}\mathrm{Zn}}}
\newcommand{\ye}{\ensuremath{Y_\mathrm{e}}}
\newcommand{\msun}{\ensuremath{M_\sun}}
\newcommand{\mini}{\ensuremath{M_\mathrm{ini}}}
\newcommand{\mej}{\ensuremath{M_\mathrm{ej}}}
\newcommand{\mch}{\ensuremath{M_\mathrm{Ch}}}
\newcommand\ddfrac[2]{\frac{\displaystyle #1}{\displaystyle #2}}

\begin{document}

\title{Remnants and ejecta of thermonuclear electron-capture supernovae}
\subtitle{Constraining oxygen-neon deflagrations in high-density white dwarfs}

\author{
	S.~Jones\inst{1,12}
	\and
	F.~K.~R\"opke\inst{2,3}
	\and
	C.~Fryer\inst{1}
	\and
	A.~J.~Ruiter\inst{4}
	\and
	I.~R.~Seitenzahl\inst{4}
        \and
	L.~R.~Nittler\inst{5}
	\and
        S.~T.~Ohlmann\inst{6}
	\and
	R.~Reifarth\inst{7,12}
	\and
	M.~Pignatari\inst{8,9,10,12}
	\and
	K.~Belczynski\inst{11}
}

\institute{
	X Computational Physics (XCP) Division, Los Alamos National Laboratory, New Mexico 87544, USA\\
	\email{swjones@lanl.gov}
	\and
	Heidelberg Institute for Theoretical Studies, Schloss-Wolfsbrunnenweg 35, 69118 Heidelberg, Germany
	\and
	Zentrum f\"ur Astronomie der Universit\"at Heidelberg, Albert-Ueberle-Str. 2, 69120
	Heidelberg, Germany
	\and
	School of Physical, Environmental and Mathematical Sciences, University of New South
	Wales, Australian Defence Force Academy, Canberra, ACT 2600, Australia
	\and
	Department of Terrestrial Magnetism, Carnegie Institution of Washington, 5241 Broad
	Branch Road NW, Washington, DC 20015, USA
	\and
	Max Planck Computing and Data Facility, Gie{\ss}enbachstraße 2, 85748 Garching, Germany
	\and
	Goethe-Universit\"at Frankfurt, 60438, Frankfurt a.M., Germany
	\and
	E. A. Milne Centre for Astrophysics, Department of Physics \& Mathematics, University of Hull, HU6 7RX, UK
	\and
	Joint Institute for Nuclear Astrophysics - Center for the Evolution of the Elements, USA
	\and
	Konkoly Observatory, Research Centre for Astronomy and Earth Sciences, Hungarian Academy of Sciences, Konkoly Thege Miklos ut 15-17, H-1121 Budapest, Hungary
	\and
	Nicolaus Copernicus Astronomical Center, Polish Academy of Sciences, Bartycka 18, PL-00-716 Warsaw, Poland
	\and 
	NuGrid Collaboration, \url{http://nugridstars.org}
}

\date{Received October 9, 2018; accepted December 13, 2018}

\abstract
{
	The explosion mechanism of electron-capture supernovae (ECSNe) remains equivocal: it
	is not completely clear whether these events are implosions in which neutron stars are
	formed, or incomplete thermonuclear explosions that leave behind bound ONeFe white
	dwarf remnants. Furthermore, the frequency of occurrence of ECSNe is not known,
	though it has been estimated to be of the order of a few per cent of all core-collapse
	supernovae.
	We attempt to constrain the explosion mechanism (neutron-star-forming implosion or
	thermonuclear explosion) and the frequency of occurrence of ECSNe using nucleosynthesis
	simulations of the latter scenario, population synthesis, the solar abundance
	distribution, pre-solar meteoritic oxide grain isotopic ratio measurements and the
	white dwarf mass-radius relation.
	Tracer particles from 3d hydrodynamic simulations were post-processed
	with a large nuclear reaction network in order to determine the complete
	compositional state of the bound ONeFe remnant and the ejecta, and
	population synthesis simulations were performed in order to estimate the
	ECSN rate with respect to the CCSN rate.
	The 3d deflagration simulations drastically overproduce the neutron-rich isotopes
	\Ca{48}, \Ti{50}, \Cr{54}~, \Fe{60} and several of the Zn isotopes relative to their
	solar abundances. Using the solar abundance distribution as our constraint, we place an
	upper limit on the frequency of thermonuclear ECSNe as 1$-$3~\% the frequency at which
	core-collapse supernovae (FeCCSNe) occur. This is on par with or 1~dex lower than the
	estimates for ECSNe from single stars. The upper limit from the yields is also in
	relatively good agreement with the predictions from our population synthesis
	simulations. The \Cr{54}/\Cr{52}~and \Ti{50}/\Ti{48} isotopic ratios in the ejecta are
	a near-perfect match with recent measurements of extreme pre-solar meteoritc oxide
	grains, and \Cr{53}/\Cr{52} can also be matched if the ejecta condenses before mixing
	with the interstellar medium.
	The composition of the ejecta of our simulations implies that ECSNe, including
	accretion-induced collapse of oxygen-neon white dwarfs, could actually be partial
	thermonuclear explosions and not implosions that form neutron stars.  There is still
	much work to do to improve the hydrodynamic simulations of such phenomena, but it is
	encouraging that our results are consistent with the predictions from stellar evolution
	modelling and population synthesis simulations, and can explain several key
	isotopic ratios in a sub-set of pre-solar oxide meteoritic grains. Theoretical
	mass-radius relations for the bound ONeFe WD remnants of these explosions are
	apparently consistent with several observational WD candidates. The composition of the
	remnants in our simulations can reproduce several, but not all, of the
	spectroscopically-determined elemental abundances from one such candidate WD.
}


\maketitle
%

\section{Introduction}

The fate of stars with initial masses between approximately 8 and 10~\msun~is believed to be
either an oxygen-neon (ONe) white dwarf (WD) and a planetary nebula, or a neutron star (NS) and a
supernova (SN) remnant \citep[see][for a recent review]{Doherty2017a}. In the latter case, the
event is called an electron-capture supernova \citep[ECSN;][]{Miyaji1980,Nomoto1987}.

Electron-capture supernovae are instigated by the electron capture sequence
$^{20}\mathrm{Ne}\rightarrow{^{20}\mathrm{F}}\rightarrow{^{20}\mathrm{O}}$ in
degenerate ONe stellar cores or ONe WDs that reach the Chandrasekhar limit
(\mch). If the progenitor is an isolated star it will consist of a
degenerate ONe core inside an extended H envelope, and the core will have grown
to \mch~via many recurrent thermally unstable He shell burning episodes
\citep{Ritossa1999,Jones2013}. If the progenitor star was born in a close binary
system its envelope can be stripped following the main sequence and the core can
grow to \mch~via stable He shell burning \citep{Podsiadlowski2004a,Tauris2015a}.
Finally, the progenitor could also be an ONe WD stably accreting from a binary
companion, retaining enough mass for the WD to reach \mch~\citet{Schwab2015a}.

The $\gamma$-decay of $^{20}$O heats the surrounding
plasma and results in the ignition of Ne and O burning, which proceeds in a thermonuclear
runaway because of the degenerate nature of the plasma. The burning moves outwards in a
conduction front \citep{Timmes1992} behind which the electron densities are very large and so
the ashes of the burning deleptonize quickly.  The fate of the object depends upon whether the
energy release from the nuclear burning can lift the degeneracy and blow up the star
\citep{Nomoto1991,Isern1991,Canal1992,Jones2016a,Leung2017a}, or whether the deleptonization is
so rapid that the star can never recover through nuclear burning and collapses into a neutron
star \citep{Miyaji1980,Miyaji1987,Nomoto1987,Kitaura2006,Fischer2010,Jones2016a}. We
distinguish these two fates semantically as explosion vs implosion, or tECSN (thermonuclear
ECSN) vs cECSN (collapsing ECSN)\footnote{It is worth mentioning that both tECSNe and cECSNe
are expected to be fainter than ``normal'' SNe, and therefore it is perhaps tenuous to label
such events as SNe at all.}. It is currently believed that the ignition of burning due to
electron captures on $^{20}$Ne results in a collapse that can not be reversed by thermonuclear
burning, resulting in implosion and a cECSNe. Which outcome is realized depends on the central
density of the star when the deflagration wave is ignited, which depends intimately on the
strength of the ground state--ground state second forbidden transition from $^{20}$Ne to
$^{20}$F \citep{Martinez-Pinedo2014a,Schwab2015a,Jones2016a}, which has now been measured
\citep{Kirsebom2018a}, but the impact of the new measurement remains to be fully explored.  The
precise ignition conditions have been shown by \citet{Schwab2017a} to be sensitive to the mass
fractions of Urca nuclei $^{25}$Mg and $^{23}$Na as well as to the accretion or core growth
rate.

Determining the initial stellar mass range and, hence, the frequency of occurrence of ECSNe is
a difficult undertaking. One way of doing this is by simulating both binary and single stellar
models across the initial mass range $8 \lesssim M_\mathrm{ini} / \msun \lesssim 12$, where
super-AGB stars are created, and noting the initial mass range for which ECSNe occur. Then,
assuming one knows the IMF (including for binary and triple-star systems), the correct
statistics should follow from integrating the IMF over the initial mass range for ECSNe.

\begin{figure*}
	\centering
	\includegraphics[width=0.49\textwidth]{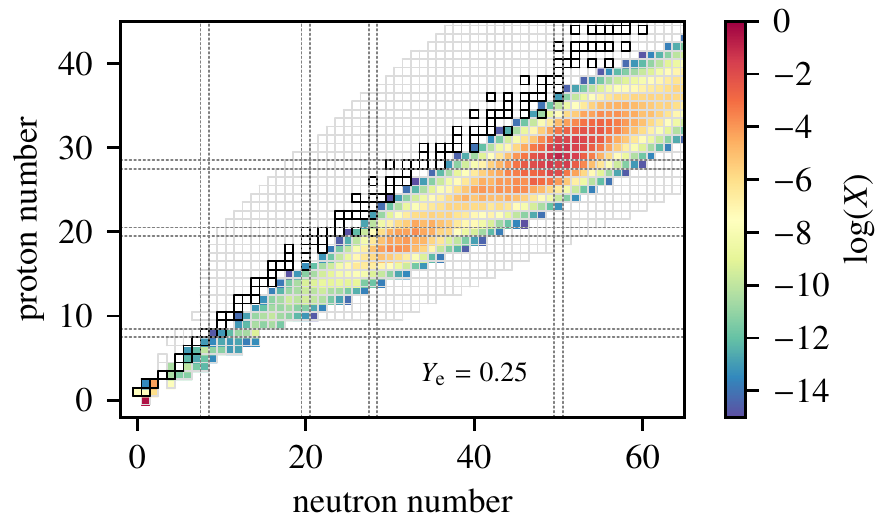}
	\includegraphics[width=0.49\textwidth]{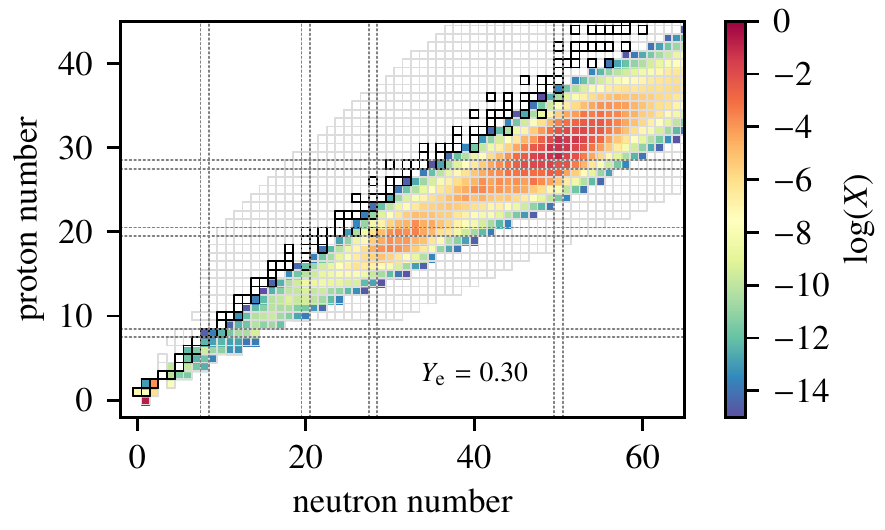} \\ 
	\includegraphics[width=0.49\textwidth]{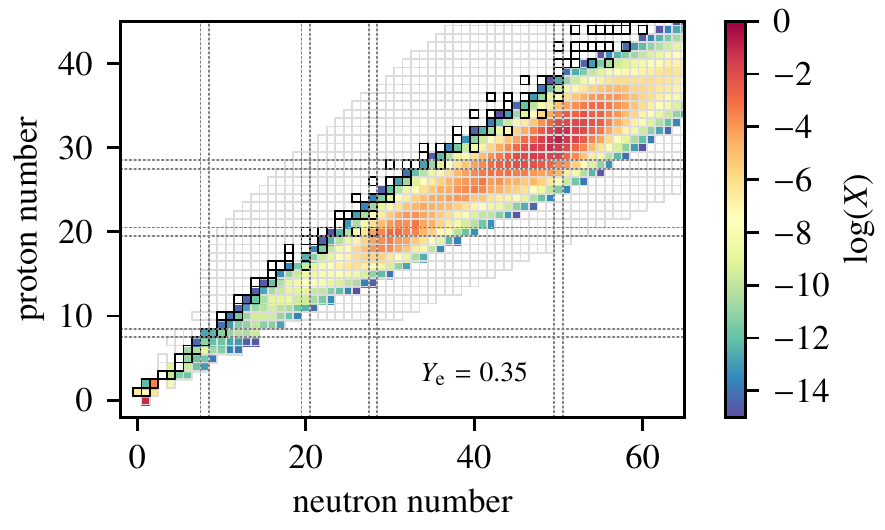}
	\includegraphics[width=0.49\textwidth]{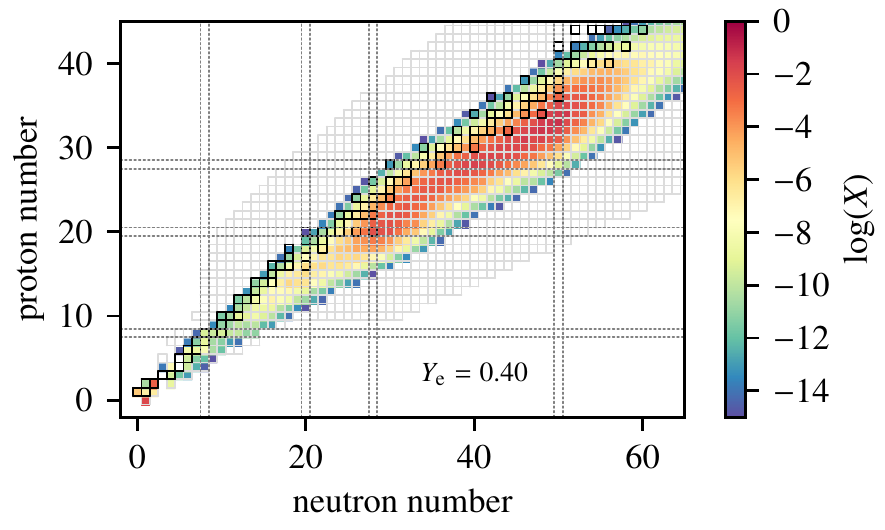} \\ 
	\includegraphics[width=0.49\textwidth]{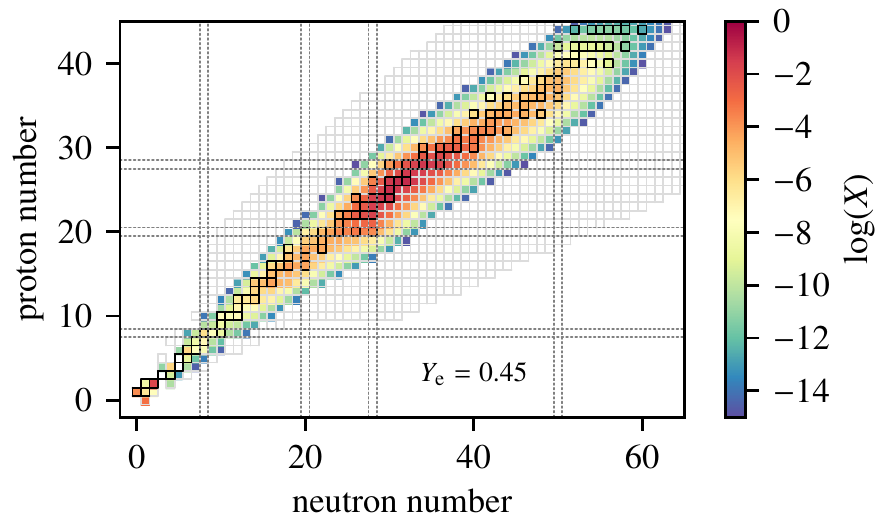}
	\includegraphics[width=0.49\textwidth]{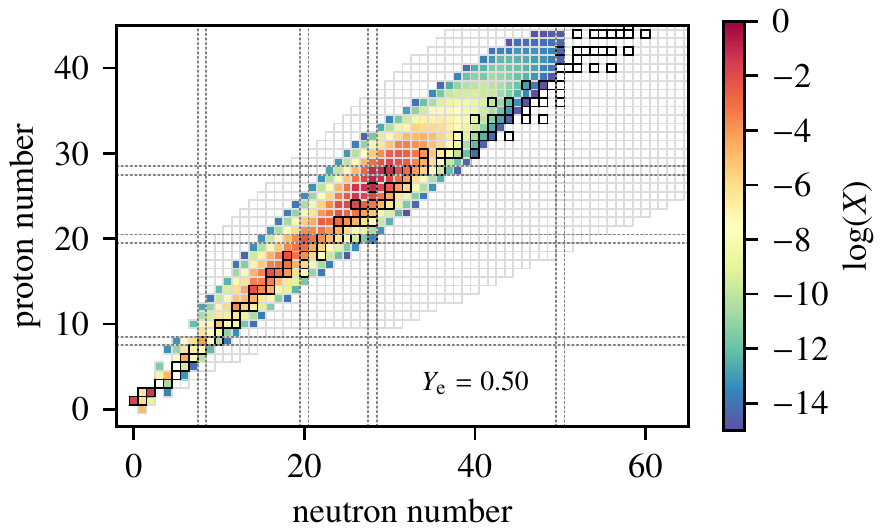} \\
	\caption{
		NSE distributions for varying electron fraction \ye~at $T=9$~GK and
		$\rho=10^{10}$~g~cm$^{-3}$. The full reaction network is outlined in grey
		squares (it actually extends up to $^{276}$Bi with $Z=83$ and $N=193$). Stable
		isotopes are outlined in thicker, black squares. The vertical and horizontal
		dotted lines mark the magic neutron and proton numbers (shell closures) at 8,
		20, 28 and 50.
	}
	\label{fig:NSE}
\end{figure*}

These stellar evolution simulations are challenging, for a number of reasons. Firstly,
super-AGB stars undergo several thousand thermal pulses (TPs;
\citealp{Ritossa1996,Siess2010a,Jones2013,Doherty2015}), in-between which the core increases
very slowly in mass, at a rate of approximately $10^{-7}-10^{-6}~\msun~\mathrm{yr}^{-1}$. The
rate of core growth depends crucially, however, on the efficiency, $\lambda$, of the third
dredge-up \citep{Herwig2012a,Ventura2013a,Jones2016b,Doherty2017a}. The thermal pulses
themselves are the result of thermal instabilities in the He-burning shell, which is of the
order of a mere $10^{-4}-10^{-5}~\msun$ of material \citep[see, e.g.][their Figure
15]{Ritossa1996}.  Furthermore, the H burning shell resides inside the lower bound of the
convective H envelope (hot bottom burning;
\citealp{Ventura2005a,Ventura2005b,Ventura2011a,Doherty2014a}).  Resolving these phenomena for
the entire evolution can require several hundreds of thousands, if not millions, of
computational time steps in the stellar evolution calculation \citep{Jones2013}.  Even then,
the physics of convective boundary mixing during the TP-SAGB (CBM; \citealp{Jones2016b}) and
the TP-SAGB wind mass loss rates are not known well enough (or modelled well enough;
\citealp{Groenewegen2018a}) to accurately predict the dredge-up efficiency or the time at which
the envelope would be completely expelled into the interstellar medium
\citep{Siess2007,Poelarends2008,Doherty2017a}.  Understanding mass loss is further complicated
by the fact that a substantial fraction of super-AGB stars are not isolated in space but exist
in binary systems \citep{Duchene2013a} in which the companions will exchange mass during their
lifetimes
\citep{Podsiadlowski2004a,Sana2012a,Tauris2015a,Tauris2017a,Poelarends2017a,Siess2018a}.  At
the upper end of this mass range, the evolution is challenging to model for different reasons.
Most or all of the burning phases following C burning (that is, Ne, O and Si burning) ignite
substantially away from the centre of the star and burn inwards as convectively bounded flames
\citep{Timmes1994,Jones2013,Woosley2015a}. These flames are typically not resolvable in stellar
evolution calculations and so they tend to either burn inwards in some fashion influenced
heavily by the numerical treatment, or they are quenched \citep[perhaps somewhat artificially,
e.g.][]{Lecoanet2016a} by mixing across them, induced by CBM from the bounding convection zone
above \citep{Jones2014a}.  Simulating the evolution of the core during these events is also
very time consuming and because the model depends on the numerical treatment, the
accuracy is limited.

Another way of constraining the initial mass range for ECSNe would be to examine the
observational statistics of their observable properties and compact remnants.  In the case of a
cECSN, the neutron stars produced are thought to have baryonic (gravitational) masses of around
1.35~\msun~(1.26~\msun; \citealp[e.g.][]{Schwab2010}).  cECSN have similarities to the collapse
produced when a white dwarf accretes sufficient matter to exceed the Chandrasekhar limit, also
known as accretion induced collapse (AIC).  Simulations of AICs predict remnant (gravitational)
masses in the range $1.1-1.3~\msun$
\citep{hillebrand84,baron87,fryer99,dessart06,abdikamalov10}.  Both cECSN and AIC should
produce similar compact remnant velocities.  Because of the steep density profile, these
systems produce explosions quickly.  A number of mechanisms have been proposed to produce
neutron star kicks \citep{fryer04}.  If the remnant kick is produced through low-mode
convection (that typically takes a longer timescale to develop), these systems will have low
kick velocities \citep{herant95,fryer04,Podsiadlowski2004a,Knigge2011}.  If, instead, the kick
is driven by asymmetries in the collapsing core, these systems will have strong kicks because
convection will not have time to wash out the asymmetric collapse \citep{burrows96,fryer04}.

Frustratingly, progenitors at the low-mass end of ``regular'' iron-core-collapse supernovae
(FeCCSNe) also have relatively steep density gradients at the edge of the core, and similar
core masses \citep{Mueller2016a}.  Therefore it could be challenging to distinguish between
neutron stars formed via cECSNe and those formed from the lower-mass end of the massive star
mass range that explode as FeCCSNe\footnote{see, however, the recent work by
	\citet{Gessner2018a}, in which ECSNe are shown to impart even lower kicks than low-mass
FeCCSNe to the nascent neutron star.}. 

Light curves of cECSNe from single stars are expected to be characterized by low peak
bolometric luminosities and low $^{56}$Ni ejecta masses compared to FeCCSNe from more massive
FeCCSN progenitors with zero-age main sequence mass $M\gtrsim12~\msun$.  A number of such
events have indeed been observed
\citep[e.g.][]{Turatto1998a,Botticella2009a,Fraser2011a,Kulkarni2007a,VanDyk2012a} and reported
as candidate cECSNe, but there are also theories that these optical transients are outbursts
from massive stars \citep[so-called supernova-impostors,
e.g.][]{Kulkarni2007a,Bond2009a,Berger2009a} and not supernovae at all, or that they are
FeCCSNe from massive stars in which there is a substantial amount of fallback, particularly of
the $^{56}$Ni \citep{Turatto1998a}. In some cases candidate cECSN optical transients have even
been ruled out as being exploding super-AGB stars \citep{Eldridge2007}.

\begin{figure*}
	\centering
	\includegraphics[width=\textwidth]{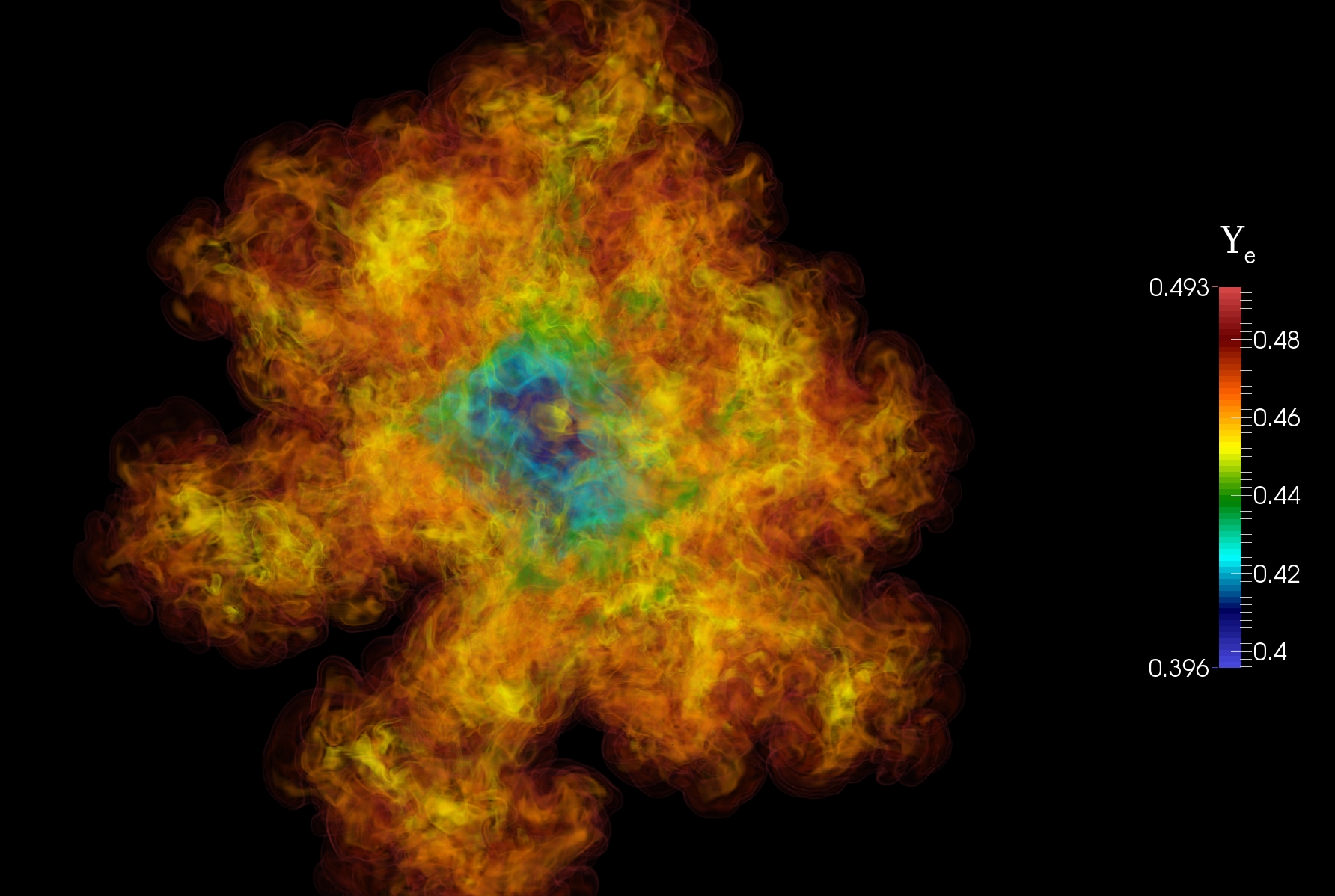}
	\caption{
		Volume rendering showing the spatial distribution of electron fraction \ye~in the
		deflagration ashes of the $512^3$ ONe deflagration simulation G14.
	}
	\label{fig:yevol}
\end{figure*}

Recently, \citet{Jerkstrand2018a} published spectral synthesis results for the lowest-mass
FeCCSN progenitor from \citet{Sukhbold2016a} computed with the KEPLER stellar evolution code
and exploded with the P-HOTB code in 1D \citep[see][for
details]{Ugliano2012,Ertl2016a,Sukhbold2016a,Jerkstrand2018a}.  \citet{Jerkstrand2018a} point
out that of the three sub-luminous IIP SNe SN 1997D, SN 2005cs and SN 2008bk, all show He and C
lines in their nebular spectra that are thought to originate from a thick He shell. This is
consistent with massive star progenitors but not with super-AGB progenitors, adding weight to
the interpretation of these three (and perhaps other) sub-luminous IIP SNe as being FeCCSNe
from low-mass massive star progenitors and not cECSNe. Another low-luminosity IIP is SN
2016bkv, which does not exhibit the He and C lines associated with the He shell, but does still
have O lines \citep[][their Figure~8]{Hosseinzadeh2018a}, would be more consistent with a cECSN
than SN 1997D, SN 2005cs or SN 2008bk, however its apparently large radioactive Ni ejecta mass
is in tension with cECSN models. It cannot be ruled out, however, that the apparent enhancement
of radioactive Ni in the ejecta of SN 2016bkv stems from an incomplete consideration of the CSM
interaction in the modelling.  There may still be life in the observational prospects of
detecting ECSNe: super-AGB stars likely have low velocity winds with high mass loss rates,
creating a dense circumstellar medium (CSM) around the star.  \citet{Moriya2014a} showed that
because of this CSM, ECSNe would be of type IIn (H-rich but with narrow spectral lines from the
slow-moving CSM), that may be consistent with the Crab supernova \citep{Smith2013a}, which has
been proposed to be the remnant of an ECSN.

\begin{figure*}
	\centering
	\includegraphics[width=.9\textwidth]{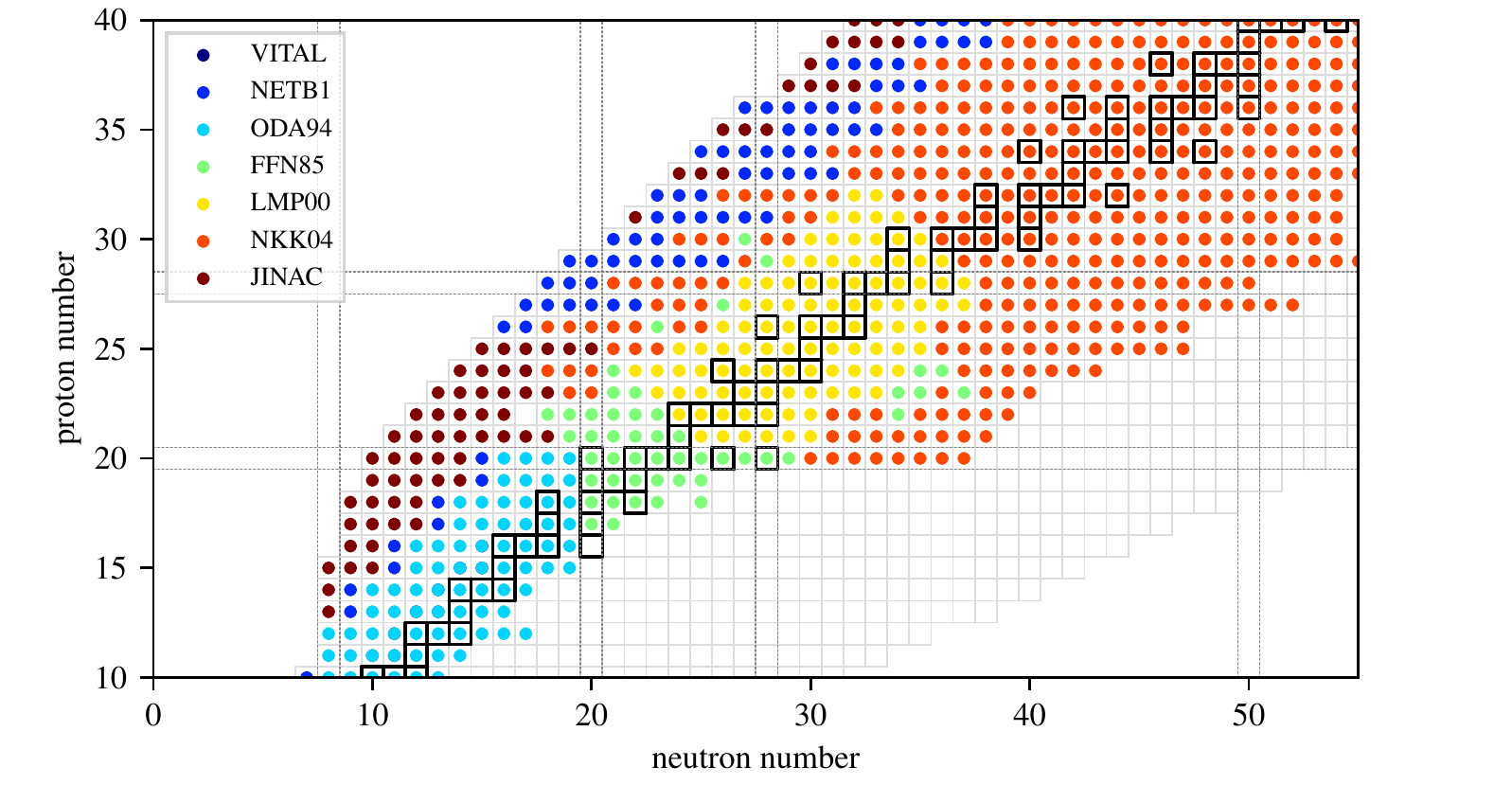} \\
	\includegraphics[width=.9\textwidth]{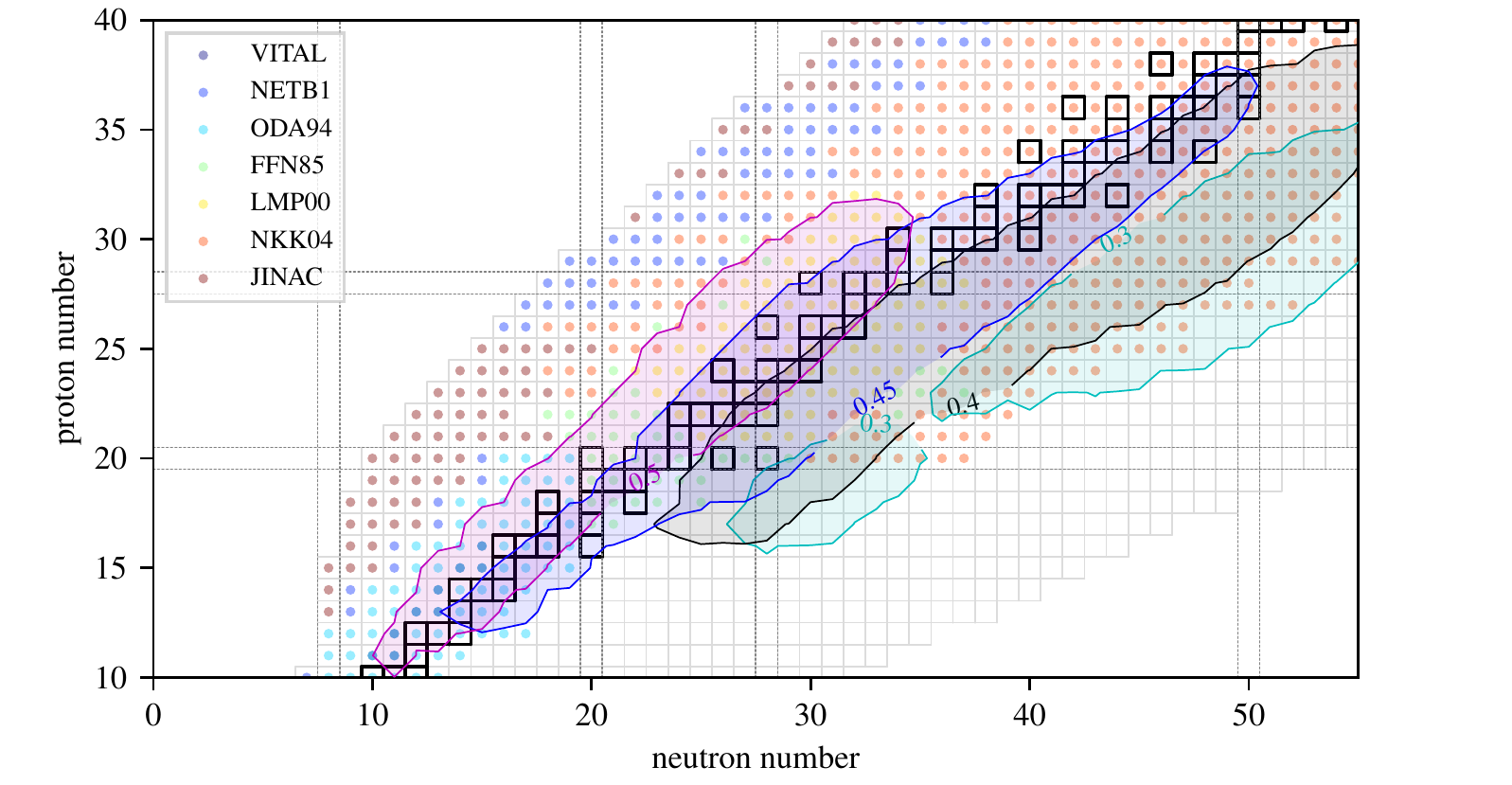}
	\caption{
		\emph{Top Panel:} Sources of electron capture and positron decay ($\beta^+$) reaction rates in
		our reaction network. The labels are as follows: VITAL (only
		$^{7}\mathrm{Be}+\mathrm{e}^-\rightarrow^{7}\mathrm{Li}+\nu+\gamma$
		from \citealp{CaughlanFowler1988});
		NETB1 (weak reaction rates from
		NetGen -- \href{http://www.astro.ulb.ac.be/Netgen}{http://www.astro.ulb.ac.be/Netgen} --
		which are predominantly from \citealp{Takahashi1987a}
		and \citealp{Goriely1999a});
		ODA94 \citep{ODA94}; FFN85
		\citep{FFNweak1985}; LMP00 \citep{Langanke2000}; NKK04 \citep{Nabi2004a}; JINAC
		\citep{Cyburt2010}.
		\emph{Bottom Panel:} Same as top panel; shaded regions cover isotopes with mass
		fraction greater than $X=5\times10^{-6}$ for an NSE state at $T=9$~GK and
		$\rho=10^{10}$~g~cm$^{-3}$. The value of \ye~for each shaded region is written
		on the enclosing contour line. One can see that at $\ye=0.45$ there are a
		substantial number of isotopes outside of the region covered by the
		\citet{Langanke2000} tables with mass fractions greater than
		$X=5\times10^{-6}$, indicating that they could potentially contribute to the
		evolution of \ye~when their weak reaction rates are considered, which is what
		we observe in our simulations when including the \citet{Nabi2004a} rates.
	}
	\label{fig:network}
\end{figure*}

If we continue down the path of not finding a transient or detecting a progenitor or a remnant
that can unambiguously be identified as a smoking gun for a cECSN, then one must draw the
conclusion that either
(1) super-AGB stars never reach the conditions for explosion in isolated systems
(i.e. not in an interacting binary);
(2) that cECSNe are even less frequent than currently predicted, or
(3) that our understanding of the explosion mechanism (and hence the synthetic
observables, such as nucleosynthesis yields and light curves, spectra) is less
complete than previously thought.
This paper is an exploration of the possibility presented in point 3.  \citet{Nomoto1991},
\citet{Isern1991}, \citet{Canal1992} and more recently \citet{Jones2016a} have suggested that
there is a possibility that ECSNe could be thermonuclear explosions (i.e. exploding tECSNe
rather than imploding cECSNe), as described above.  Instead of collapsing into a neutron star,
in a tECSN a portion of the core is ejected leaving a gravitationally bound WD remnant
consisting of O, Ne and Fe-group elements (ONeFe WD) behind. It is these explosions that are
the focus of this paper. There are two important predictions that can be used to constrain
whether or not tECSNe can occur and at what frequency: the ejected material should contribute
to galactic chemical evolution (GCE) and therefore to the solar chemical inventory, and the
bound ONeFe WD remnants should still exist within our Galaxy. In both cases there are chemical
signatures that are unique to these events owing to the extreme conditions under which the
thermonuclear burning proceeds, compared to SNe Ia.

Building on the hydrodynamic tECSN simulations already performed by \citet{Jones2016a}, we
calculate the full nucleosynthesis in the ejecta and the bound remnant in the tECSN
simulations.  We also perform binary population synthesis simulations to obtain a theoretical
estimate of the ECSN rate with respect to the FeCCSN rate should all ECSNe be tECSNe. We
examine the nucleosynthesis results in the context of GCE, placing an upper limit on the
frequency of occurrence of tECSNe by comparing to the solar abundance distribution. The
composition of the ejecta is compared with recent measurements of pre-solar meteoritic oxide
grains exhibiting extreme isotopic ratios for Cr and Ti, for which tECSNe are found to be a
remarkably good match. Lastly, we compute mass-radius relations for the bound ONeFe WD remnants
and compare them with both the population synthesis results and observational WD surveys.

It is worth re-emphasizing at this point that we believe the outcome of ECSNe -- cECSN,
implosion and NS or tECSN, explosion and ONeFe WD -- remains an open question at present. This
is because obtaining a convincing answer using simulations depends on several modelling
assumptions and microphysics constraints, as described by \citet{Jones2016a}.  Additionally,
the ignition density of the deflagration remains uncertain, which is critical input for the
hydrodynamic simulations. We hope that this study of the nucleosynthesis and compact remnants
in the case of a partial thermonuclear explosion brings us closer to an answer.

\section{Post-processing technique and reaction network}

\subsection{Nuclear reaction network: approach and methods}

The simulations presented in \citet{Jones2016a} included the advection of
$\sim10^6$ equal-mass tracer particles, as has been described in several
previous works
\citep{Travaglio2004a,Roepke2006a,Seitenzahl2010a,Seitenzahl2013a}. In this
work, we performed nucleosynthesis simulations of these tracer particles in
post-processing, taking the temperature and density evolution of the tracer
particles as a function of time and integrating the reaction equations for those
conditions. For the post-processing, a derivative of the
NuGrid\footnote{\href{http://www.nugridstars.org}{nugridstars.org}} nuclear
reaction network was used \citep[as has briefly been described
in][]{Pignatari2016a,Ritter2018b}. The network was substantially renovated and
updated to use the screening corrections for fusion reactions by
\citet{Chugunov2007a} and the semi-implicit extrapolation method by
\citet{Bader1983a} and \citet{Deuflhard1983a} was implemented \citep[see
also][]{Timmes1999a}, which was used for all simulations presented here. A new
nuclear statistical equilibrium (NSE) solver was also written largely following
the work of \citet{Seitenzahl2009a}, and the NSE state solution at a given
$(T,\rho,\ye)$ is now coupled to the weak reactions (for the time-dependence of
the electron fraction \ye) using a Cash-Karp type Runge-Kutta integrator
\citep{Cash1990a}. Reverse reaction rates were computed in real time from their
forward rates using the principle of detailed balance \citep[see the appendix
of][for a concise formulation]{Calder2007a}. This improved the agreement between
the reaction network and the NSE solver. The network dynamically adapts the
problem size at every integration step in order to minimize the computational
cost of the matrix inversion that must be performed at least twice per time step
(i.e. for the first two levels of the Bader-Deuflhard integrator with $n=2$ and
$n=6$). The matrix is written directly into a sparse format, after which it is
compressed down to the problem size for the time step and the LU decomposition
and subsequent back-substitutions are then performed using the SuperLU sparse
matrix library \citep{superlu99,superlu_ug99,li05} together with the
OpenBLAS\footnote{\href{http://www.openblas.net}{openblas.net}} BLAS library.

The abundance distributions were post-processed for a second time to account for the
radioactive decay of the ejecta following the explosion. Only spontaneous decays occur in the
cold, low density environment of the ejecta. The decay rates were assumed to be the same as
under terrestrial conditions where many experimental data exist. The decays of isotopes with
mass fractions $>10^{-20}$ were processed with a relative uncertainty of better than 1~\%.

\begin{figure}
	\centering
	\includegraphics[width=0.5\textwidth]{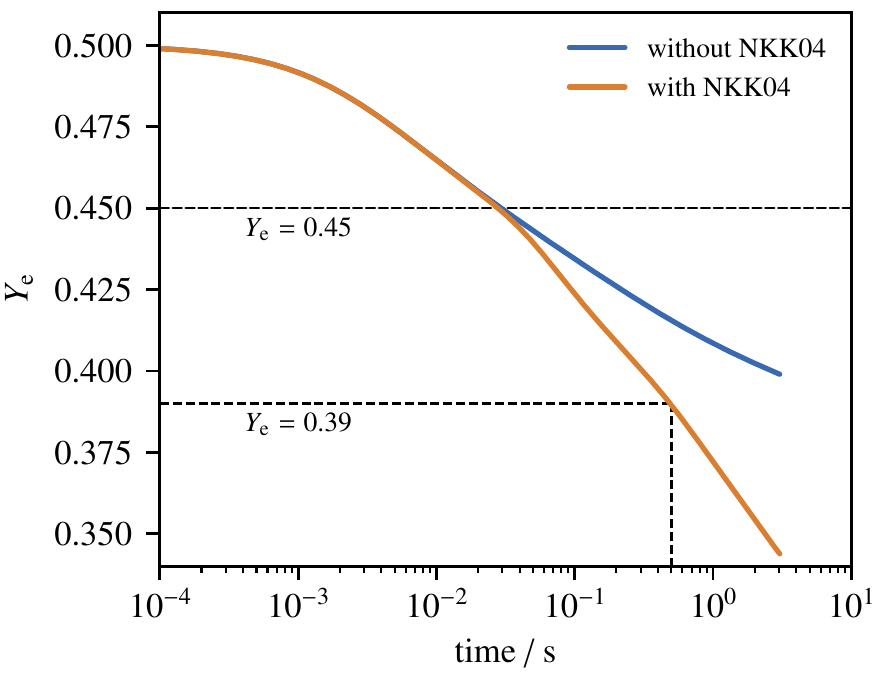}
	\caption{
		Evolution of electron fraction \ye~in a constant temperature and density
		network integration with and without the weak rates by \citet{Nabi2004a}
		included in the reaction network at their positions from
		Figure~\ref{fig:network}. The conditions were $T=9$~GK,
		$\rho=10^{10}$~g~cm$^{-3}$ and $\ye(t=0) = 0.5$. The two curves separate below
		$ye\approx0.45$, where the contribution of the \emph{fp} and \emph{fpg} shell nuclei
		to the deleptonization become significant.
	}

	\label{fig:ye_with_NKK}
\end{figure}

\subsection{Neutron-richness in thermonuclear explosions}

Before we construct a nuclear reaction network and choose a suitable set of reaction rates, we
first explore the conditions under which nucleosynthesis occurs in tECSNe.  Although the
physical mechanism is similar to thermonuclear explosions in CO white dwarfs, we find that the
reaction networks used in these studies \citep{Travaglio2004a,Roepke2005e,Seitenzahl2010a} are
insufficient for our models.

While in the context of Type Ia supernova explosion models two modes for the propagation of
thermonuclear combustion fronts are discussed -- subsonic deflagrations and supersonic
detonations \citep[see, e.g.][]{Roepke2017a} -- our models of electron capture supernovae
assume burning to proceed in the subsonic deflagration regime exclusively. The burning products
by the deflagration in high density ONe cores or WDs can generally be well described by nuclear
statistical equilibrium \citep[see, e.g.][]{Seitenzahl2009a}. That is, the timescale for the
strong reactions to equilibrate is shorter than the timescale on which the local thermodynamic
conditions of the material are changing. In this case, that is the hydrodynamic time-scale. The
weak reaction rates (electron/positron-captures and $\beta^\pm$-decays) for the prevalent
isotopes are typically much slower than the strong reactions and can not necessarily be assumed
to reach equilibrium.

The isotopic composition of material in NSE depends critically on $\ye$.  Figure~\ref{fig:NSE}
shows six pseudo-colour plots of the isotopic nuclear chart, where the colour scale represents
the mass fractions of the isotopes for NSE solutions with $T=9$~GK, $\rho =
10^{10}$~g~cm$^{-3}$  (i.e.~a typical state that is reached in a tECSN) and $\ye = \{0.25, 0.3,
0.35, 0.4, 0.45, 0.5\}$. It is relatively well-known that the NSE distribution will tend to
favour nuclei with a ratio of proton number to mass number $Z/A \approx\ye$
\citep[e.g.][]{Clifford1965a,Hartmann1985a}, which can be seen in the most abundant (red)
isotopes in Figure~\ref{fig:NSE}, which tend to more neutron-rich nuclei for lower \ye.  It is
interesting that in Figure~\ref{fig:NSE} one can clearly see the bifurcation of the
peak NSE distribution (aside from the free nucleons) at lower values of electron fractions,
with the two peaks staying close to the intersection of the magic neutron and proton numbers at
$Z = \{20,28\}$ and $N = \{28,50\}$.

The values of $\ye$ reached in models for Type Ia supernova explosions depend on the
metallicity of the progenitor star and its central density. The most massive Type Ia supernova
progenitors are postulated to be degenerate CO white dwarf stars with masses at the
Chandrasekhar limit, which for a non-rotating CO white dwarf star would be close to
$1.4~M_\odot$. It has been argued that differential rotation can substantially increase the
mass supported against collapse \citep{steinmetz1992a, pfannes2010a, pfannes2010b}  and some
observed superluminous Type Ia supernovae have been associated with explosions of progenitors
with masses above $1.4~M_\odot$ (e.g. \citealp{howell2006a}; for an overview see
\citealp{taubenberger2017a}), but burning is not expected to take place at extremely high
densities in these scenarios and the results are inconsistent with observed supernovae
\citep{fink2018a}.  One widely discussed scenario for normal Type Ia supernovae is that a
thermonuclear runaway of C in Chandrasekhar-mass WDs initiates a deflagration wave that almost
immediately enters the turbulent burning regime and later may or may not transition into a
detonation \citep[delayed detonation model, ][]{Khoklov1991a}. While a pure turbulent
deflagration is a successful model for the subluminous class of SN 2002cx-like Type Ia
supernovae \citep[e.g.][]{kromer2013a}, three-dimensional simulations of the delayed detonation
scenario reproduce many of the observational characteristics of Type~Ia supernovae
\citep[e.g.][]{Kasen2009a,Blondin2013a}, but fail in some (important) aspects \citep{Sim2013a}.
The highest density that can be achieved in a Chandrasekhar-mass Type~Ia supernova explosion is
the initial central density of the progenitor, which is about
$\rho_\mathrm{c}\approx2-5\times10^9$~g~cm$^{-3}$ for an appropriate value of the electron
fraction in the CO white dwarf \citep{Lesaffre2006a}, depending on its cooling history. At
these densities, the deflagration ashes will be buoyant (high Atwood number), resulting in
substantial expansion of the CO white dwarf of the order of a few hundred milliseconds,
together with a corresponding decrease in the maximum density. On the time-scale of a few
hundred milliseconds the deleptonization of the densest material in a CO white dwarf proceeds
only at a moderate rate and \ye~typically does not fall below about 0.46
\citep[e.g.][]{Travaglio2004a}.

Conversely, in the ONe deflagration during an ECSN the Atwood number is
substantially lower than in a Type~Ia SN owing to the higher densities and the
higher degree of electron degeneracy. The densest regions expand more slowly and
spend more time at densities where the rate of deleptonization is much faster.
If the deleptonization is fast enough and the Atwood number is low enough, the
ONe WD or the degenerate ONe core will experience a rapid decrease in \ye~and
the core will eventually collapse into a neutron star
(\citealp{Miyaji1980,Nomoto1987}; model H01 of \citealp{Jones2016a}). In less
extreme cases (i.e. lower central densities of the ONe core at the time of
deflagration ignition; $\rho_c\lesssim 10^{10}$~g~cm$^{-3}$), the nuclear energy
released may compete with the deleptonization and the buoyant acceleration of
the hot ashes can lead to the unbinding of a substantial fraction of the core.
For such a case, the minimum \ye~found in simulations is $\ye\sim0.38$
\citep[Figure~\ref{fig:yevol} and][their Figure~2]{Jones2016a}.

\begin{table*}
	\caption{
		Isotopes included in the post-processing reaction network.
	}
	\label{table:network}
	\centering
	\begin{tabular}{c c c}
		\toprule\toprule
		element & min.~$A$ & max.~$A$ \\
		\midrule
		n &     1 &     1 \\ 
 H &     1 &     3 \\ 
He &     3 &     6 \\ 
Li &     7 &     9 \\ 
Be &     7 &    12 \\ 
 B &     8 &    14 \\ 
 C &    11 &    18 \\ 
 N &    11 &    21 \\ 
 O &    13 &    22 \\ 
 F &    17 &    26 \\ 
Ne &    17 &    41 \\ 
Na &    19 &    44 \\ 
Mg &    20 &    47 \\ 
Al &    21 &    51 \\ 
Si &    22 &    54 \\ 
 P &    23 &    57 \\ 
 S &    25 &    60 \\ 
Cl &    26 &    63 \\ 
Ar &    27 &    67 \\ 
 K &    29 &    70 \\ 
Ca &    30 &    73 \\ 

		\bottomrule\bottomrule
	\end{tabular}
	\hfill
	\begin{tabular}{c c c}
		\toprule\toprule
		element & min.~$A$ & max.~$A$ \\
		\midrule
		Sc &    32 &    76 \\ 
Ti &    34 &    80 \\ 
 V &    36 &    83 \\ 
Cr &    38 &    86 \\ 
Mn &    40 &    89 \\ 
Fe &    42 &    92 \\ 
Co &    44 &    96 \\ 
Ni &    46 &    99 \\ 
Cu &    48 &   102 \\ 
Zn &    51 &   105 \\ 
Ga &    53 &   108 \\ 
Ge &    55 &   112 \\ 
As &    57 &   115 \\ 
Se &    59 &   118 \\ 
Br &    61 &   121 \\ 
Kr &    63 &   124 \\ 
Rb &    66 &   128 \\ 
Sr &    68 &   131 \\ 
 Y &    70 &   134 \\ 
Zr &    72 &   137 \\ 
Nb &    74 &   140 \\ 

		\bottomrule\bottomrule
	\end{tabular}
	\hfill
	\begin{tabular}{c c c}
		\toprule\toprule
		element & min.~$A$ & max.~$A$ \\
		\midrule
		Mo &    77 &   144 \\ 
Tc &    79 &   147 \\ 
Ru &    81 &   150 \\ 
Rh &    83 &   153 \\ 
Pd &    86 &   156 \\ 
Ag &    88 &   160 \\ 
Cd &    90 &   163 \\ 
In &    92 &   166 \\ 
Sn &    94 &   169 \\ 
Sb &    97 &   172 \\ 
Te &    99 &   176 \\ 
 I &   101 &   179 \\ 
Xe &   103 &   182 \\ 
Cs &   106 &   185 \\ 
Ba &   108 &   189 \\ 
La &   110 &   192 \\ 
Ce &   113 &   195 \\ 
Pr &   115 &   198 \\ 
Nd &   118 &   201 \\ 
Pm &   120 &   205 \\ 
Sm &   123 &   208 \\ 

		\bottomrule\bottomrule
	\end{tabular}
	\hfill
	\begin{tabular}{c c c}
		\toprule\toprule
		element & min.~$A$ & max.~$A$ \\
		\midrule
		Eu &   125 &   211 \\ 
Gd &   128 &   214 \\ 
Tb &   130 &   218 \\ 
Dy &   133 &   221 \\ 
Ho &   136 &   224 \\ 
Er &   138 &   227 \\ 
Tm &   141 &   230 \\ 
Yb &   143 &   234 \\ 
Lu &   146 &   237 \\ 
Hf &   149 &   240 \\ 
Ta &   151 &   243 \\ 
 W &   154 &   247 \\ 
Re &   156 &   250 \\ 
Os &   159 &   253 \\ 
Ir &   162 &   256 \\ 
Pt &   165 &   260 \\ 
Au &   167 &   263 \\ 
Hg &   170 &   266 \\ 
Tl &   173 &   269 \\ 
Pb &   175 &   273 \\ 
Bi &   178 &   276 \\ 

		\bottomrule\bottomrule
	\end{tabular}
\end{table*}

\subsection{Nuclear reaction network: species and rates}

Owing to the extreme conditions encountered in our models, we have to extend the nuclear
reaction network beyond the isotopes and rates usually accounted for in post-processing
thermonuclear supernova explosion models. For these simulations, we simply used the largest
pool of nuclei available in our reaction network, which is 5234. The bounds of the network on
the neutron- and proton-rich sides are determined by comparing the $\beta^\pm$-decay half lives
of the isotopes with a user-defined minimum characteristic time for the problem at hand. The
network is closed at the boundaries by ``ghost'' isotopes that are forced to instantaneously
$\beta^\pm$-decay (depending upon whether they are proton-rich or neutron-rich). For our
problem we set the minimum characteristic time to $10^{-5}$~seconds. After establishing the
boundaries of the network from this time-scale, we are left with a total of 5213 isotopes in
the network proper (see Table~\ref{table:network}).  This is almost certainly too large a
network for the problem at hand, however as one can see in Figure~\ref{fig:NSE} -- in which the
isotopes included in the network are drawn in grey squares -- in NSE at the lowest \ye~(0.25),
there are moderately abundant isotopes only a handful of neutrons away from the edge of the
network on the neutron-rich side. Similarly, at $\ye=0.5$, there are moderately abundant
isotopes only a handful of protons away from the edge of the network on the proton-rich side.
One can also see in Figure~\ref{fig:NSE} that in NSE at lower \ye~there is more material with
higher $A$, necessitating that the network extend well above $A=100$. In order not to
artificially influence our results by hitting the network boundaries, we did not attempt to
make the network any smaller, although there are ways in which this could have been done.  From
a practical standpoint, the motivation to reduce the network size originates from a desire to
also reduce the computational cost of the simulation.  However, the nuclear reaction network is
designed to perform the time-integration at each time step only for a sub-set of isotopes whose
abundances are actually changing. This means that all of the matrix inversions and
back-substitutions are much cheaper than if we were to perform them for the complete set of
5213 isotopes every time step. In order to determine which isotopes should be included in the
solve each time step, we do still need to evaluate all of the reaction rates for all of the
isotopes in the network, which does come with an additional and perhaps somewhat avoidable
computational cost.

The reaction rates in the network were taken from JINA Reaclib \citep{Cyburt2010}, KaDoNiS
\citep{Dillmann2006a}, NACRE \citep{AnguloNACRE1999} and NON-SMOKER \citep{Rauscher2000a}, as
well as from \citet{FFNweak1985}, \citet{Takahashi1987a}, \citet{Goriely1999a},
\citet{Langanke2000}, \citet{Iliadis2001a} and \citet{ODA94}.  There are also a handful of
reactions whose rates have been individually selected from the literature, including from
\citet{CaughlanFowler1988} for several reactions, \citet{Jaeger2001a} for the
$^{22}\mathrm{Ne}(\alpha,n)^{25}\mathrm{Mg}$ reaction, \citet{Imbriani2005a} for the
$^{14}\mathrm{N}(p,\gamma)^{15}\mathrm{O}$ reaction, several proton capture reactions from
\citet{Champagne1992a}, \citet{Fynbo_3a_2005} for
$^{4}\mathrm{He}(2\alpha,\gamma)^{12}\mathrm{C}$, \citet{Kunz2002} for
$^{12}\mathrm{C}(\alpha,\gamma)^{16}\mathrm{O}$, \citet{Heil2008a} for
$^{13}\mathrm{C}(\alpha,n)^{16}\mathrm{O}$ and \citet{Rauscher1994a} for
$^{17}\mathrm{O}(n,\alpha)^{14}\mathrm{C}$. Several of the $(n,\gamma)$ reactions have been
updated from the KaDoNis release and have been listed in \citet[][footnote
13]{Denissenkov2018a}.  We also made use of the NUDAT Nuclear data files
provided by the National Nuclear Data Center (NNDC; \citealp{Kinsey1996a}). Nuclear
masses and partition funtions are as provided by the JINA Reaclib database, and are used in the
NSE solver and for calculating reverse reaction rates.

Given that the majority of the burning takes place under conditions where assuming NSE is
appropriate, we paid special care to the weak reaction rates. The top panel of
Figure~\ref{fig:network} shows the sources of electron-capture and $\beta^+$-decay rates that
we use in the reaction network. The bottom panel of Figure~\ref{fig:network} shows the same
information as the top panel but has a portion of the information about the NSE distributions
from Figure~\ref{fig:NSE} overlaid. More specifically, it shows shaded contours enclosing
regions of the isotopic chart where the isotopic mass fractions are greater than
$5\times10^{-6}$ in an NSE state for $T=9$~GK, $\rho=10^{10}$g~cm$^{-3}$ and
$\ye=\{0.3,0.4,0.45,0.5\}$. Already at $\ye=0.45$ (approximately the minimum \ye~reached in
standard type~Ia SNe) there are isotopes with relatively large mass fractions that lie outside
of the \emph{pf} shell and are therefore quite inaccessible to nuclear shell-model codes.
Nevertheless, there is a possibility that these isotopes can collectively contribute to the
overall rate of (de)leptonization in the star. Weak reaction rate tables for a large pool of
nuclei were computed and made available by \citet{Juodagalvis2010}, however they included only
electron-capture and $\beta^+$-decay reactions and did not include their inverses. This
omission is likely inconsequential if modelling FeCCSNe, however having an
as-accurate-as-possible balance of the forward and reverse rates is necessary for ECSNe when
one is attempting to determine whether the situation resolves in core collapse or not.
We have opted to use the reaction rates from the quasiparticle random phase approximation
(QRPA) calulations by \citet{Nabi2004a} for \emph{fp} and \emph{fpg} shell nuclei because of their
extensive coverage of the isotopic chart and the fact that reaction rates have been computed
for both directions. Even so, at $\ye=\{0.3,0.4\}$ there are still a handful of isotopes with
mass fractions greater than $5\times10^{-6}$ (i.e. within the shaded regions in
Figure~\ref{fig:network}, bottom panel) for which we do not have electron-capture or
$\beta^+$-decay reaction rates.

The impact of the additional weak reaction rates by \citet{Nabi2004a} is shown in
Figure~\ref{fig:ye_with_NKK} for a network integration at constant temperature ($9$~GK) and
density ($10^{10}$~g~cm$^{-3}$) and initial $\ye(t=0)=0.5$. The \ye~evolution with and without
the NKK reaction rates clearly diverge below $\ye\approx0.45$, as discussed earlier in this
Section. We also mentioned that the minimum \ye~achieved in the sub-set of partially exploding
(i.e. not collapsing) simulations by \citet{Jones2016a} was 0.39. This \ye~has been marked on
Figure~\ref{fig:ye_with_NKK} in a manner illustrating that this \ye~is reached after
0.5~seconds under these conditions. This is intuitive because the dynamical time-scale of an
ONe white dwarf with a central density of $10^{10}$~g~cm$^{-3}$ is indeed about 0.5~seconds.

\begin{table*}
	\caption{
		Relevant properties of the hydrodynamic simulations from \citet{Jones2016a}.
		Only the models that are tECSNe are shown. In order
		from left to right, the columns are: model id, grid resolution, Coulomb
		corrections included in EoS (Y/N), central density at ignition of the
		deflagration, bound ONeFe remnant mass, mass of the ejected material, average
		electron fraction in the ONeFe
		remnant, and mass fraction of iron-group elements in the ONeFe remnant.
		The model J07 is a new addition. Nucleosynthesis
		simulations have been performed for models G14 and J07, which were used for this work.
	}
	\label{table:wdmr-sims}
	\centering
	\begin{tabular}{c c c c c c c c}
		\toprule\toprule
		id & res & CCs & $\log_{10}\rho_\mathrm{c}^{ini}$ & $M_\mathrm{rem}$ &
		$M_\mathrm{ej}$ & $\langle Y_\mathrm{e}\rangle$
		& $X_\mathrm{IGE}$\\
		& & & (g cm$^{-3}$) & ($M_\odot$) & ($M_\odot$) & \\
		\midrule
		G13 & $256^3$ & N & 9.90 & 0.647 & 0.741 & 0.491 & 0.267 \\
		G14 & $512^3$ & N & 9.90 & 0.438 & 0.951 & 0.491 & 0.263 \\
		G15 & $256^3$ & Y & 9.90 & 1.212 & 0.177 & 0.493 & 0.184 \\
		J01 & $256^3$ & N & 9.95 & 0.631 & 0.768 & 0.491 & 0.271 \\
		J02 & $256^3$ & Y & 9.95 & 1.291 & 0.104 & 0.493 & 0.175 \\
		J07 & $576^3$ & N & 9.95 & 0.366 & 1.027 & 0.489 & 0.293 \\
		\bottomrule\bottomrule
	\end{tabular}
\end{table*}

\subsection{Input from hydrodynamic simulations}

In \citet{Jones2016a}, we performed 3d hydrodynamic simulations of deflagration fronts in ONe
WDs with a range of plausible ignition densities (we will use the terminology ONe deflagrations
to describe this scenario). Within the set of six models, there were five models that resulted
in a partial ejection (gravitational unbinding) of material, leaving behind a gravitationally
bound ONeFe WD (tECSNe). One model -- with the highest central density at ignition,
$2\times10^{10}$~g~cm$^{-3}$ -- collapsed into a neutron star. Of the five models that were
tECSNe, two included a correction to the internal energy and pressure in the equation of state
(EoS) from the non-ideal behaviour of the plasma (Coulomb corrections; CCs).

The relevant results from \citet{Jones2016a} are summarized in
Table~\ref{table:wdmr-sims} for convenience. This paper is concerned only with
the nucleosynthesis in the tECSNe and therefore the model H01 has been
	omitted. We note that we have added a new hydrodynamic simulation J07 to
	this work, which is a higher-resolution version of J01 from
	\citet{Jones2016a}. We added this model because we would like to compare
	the nucleosynthesis in models with different ignition densities at a
	similar (and as high as possible) numerical resolution. For this work we
will therefore be using simulations G14 and J07.
 To re-state a pertinent point from \citet{Jones2016a}: although our simulations
 do not yet exhibit convergence upon grid refinement, increasing the grid
 resolution yields a higher ejected mass, suggesting that further increasing the
 grid resolution will likely keep the outcome as a tECSN and not a
 core-collapse. Nevertheless, we admit that there is still much to do to improve
 the status of the hydrodynamic simulations, and this is currently a work in
 progress.

In each of the hydrodynamic simulations, a set of $\sim10^6$ Lagrangian tracer
particles were passively advected with the flow, sampling their local
thermodynamic environment. As a result, we obtain a trajectory for each
particle, which contains temperature and density as a function of time.
The method for assigning the tracer particle masses and initial spatial
distribution was the same as in \citet{Seitenzahl2010a}, which we briefly
summarize here for convenience, but to which we refer the interested reader for
complete details. The tracer particle masses vary smoothly with initial radius
and their distribution is broken into three spatial parts. Within some radius
$R_1$ the tracer particles have equal mass and resolve the region where the
density profile is relatively flat and most of the NSE burning takes place. For
initial radii $R_1 < R < R_2$, where the density gradient is steeper and the
density is lower, the tracer particles have equal volume and therefore provide
better sampling of the lower density material where incomplete burning
synthesizes intermediate-mass elements (IMEs). Lastly, the particles in the
exterior layer with initial radii $R > R_2$ have equal mass again. To compute
the nucleosynthesis, the particle trajectories were fed directly into the
post-processing network for each particle at the time when the deflagration
front arrives at that particle's location.

\section{Nucleosynthesis results}
\label{sec:yields}

\begin{figure*}
	\centering
	\includegraphics[width=\textwidth]{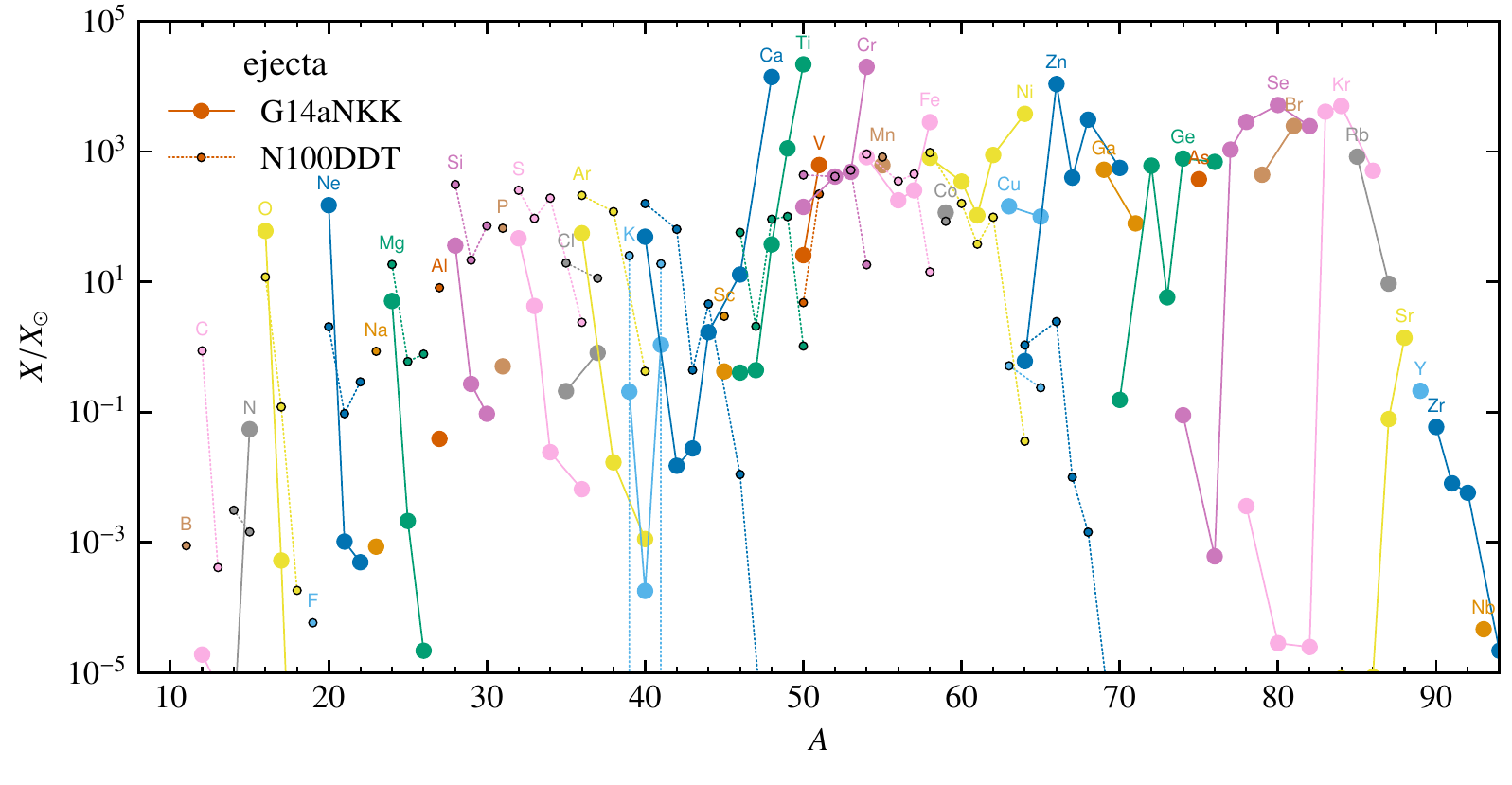}
	\caption{
	Overabundance (mass fraction relative to solar; $X/X_\odot$) of stable isotopes in the
	ejecta of simulation G14aNKK after decaying for $10^{16}$~s (0.32~Gyr). The striking production of
	\Ca{48}, \Ti{50} and \Cr{54} is a partcular hallmark of deflagrations in degenerate
	media high-density \citep[see, e.g.][]{Woosley1997a}. The DDT simulation N100DDT from
	\citet{Seitenzahl2013a} is plotted for comparison, decayed to 2~Gyr.
	}
	\label{fig:G14a_n100_nkk}
\end{figure*}

\begin{figure*}
	\centering
	\includegraphics[width=\textwidth]{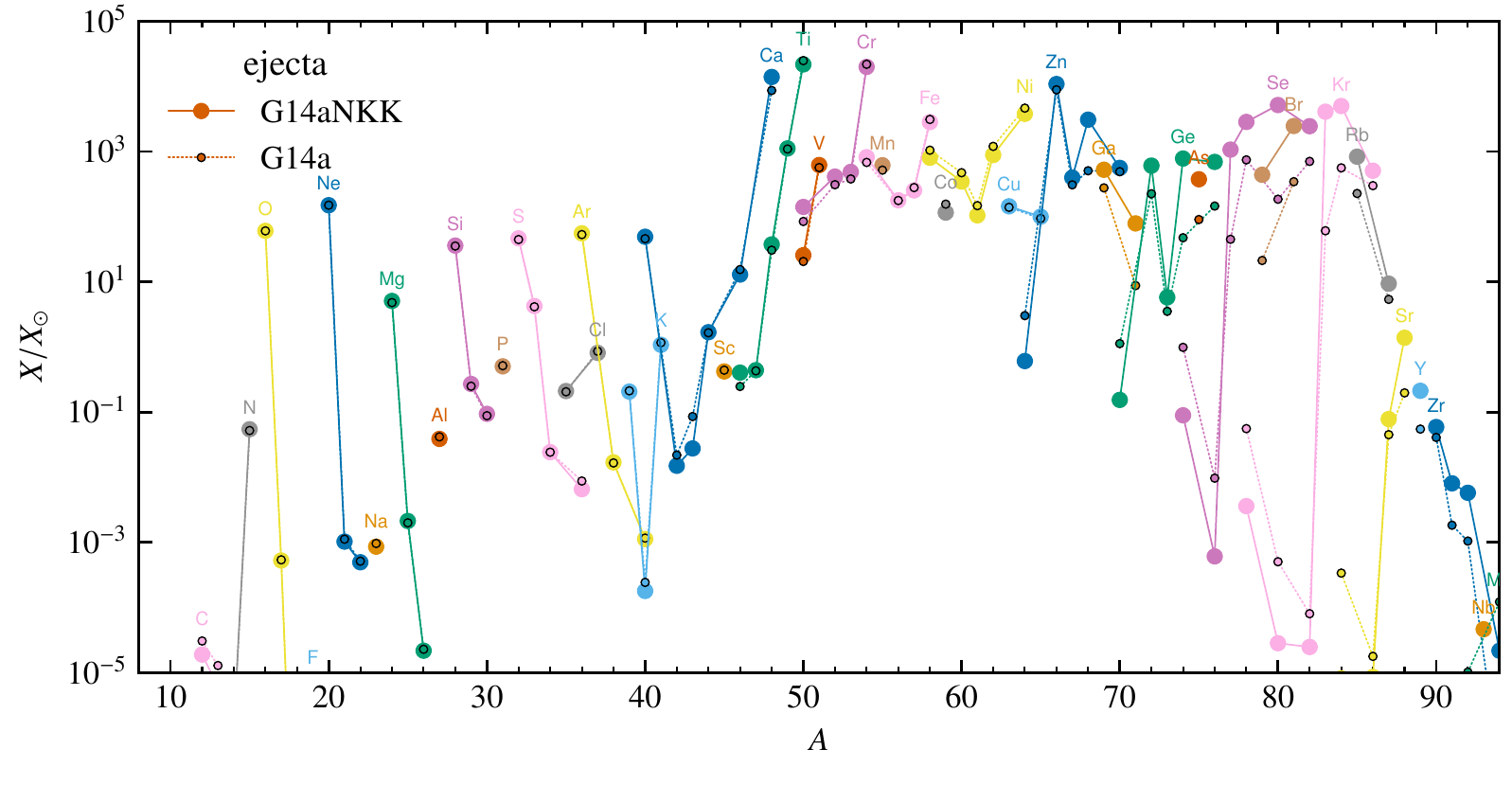}
	\caption{
	Overabundance (mass fraction relative to solar; $X/X_\odot$) of stable isotopes in the
	ejecta of simulation G14a after decaying for $10^{16}$~s. The two lines correspond to
	post-processing nucleosynthesis simulations where \citet[][NKK]{Nabi2004a} weak
	reaction rates were included ("NKK") and when they were not. Including the NKK rates
	results in a larger yield of \Ca{48} and smaller yields of \Ti{50} and \Cr{54}. The
	yields of the trans-iron elements from Ga to Nb are substantially increased when the
	NKK rates are included.
	}
	\label{fig:NKK_yield_impact}
\end{figure*}

\begin{figure*}
	\centering
	\includegraphics[width=\textwidth]{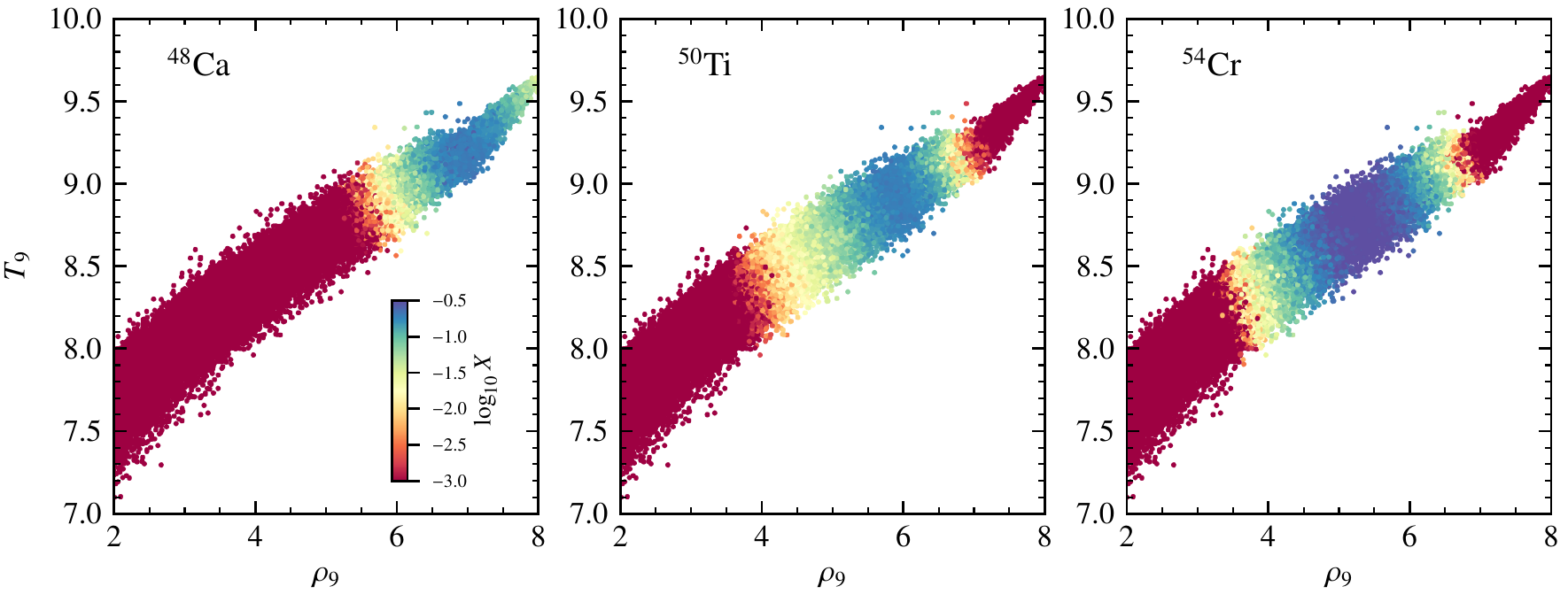}
	\caption{
		Mass fractions of \Ca{48}, \Ti{50} and \Cr{54} as a function of peak
		temperature (GK) and density ($\rho_9 = 10^9$~g~cm$^{-3}$) for
		the sub-set of
		tracer particles where these isotopes are made in abundance.
	}
	\label{fig:tps}
\end{figure*}

The nucleosynthesis in the ejecta of our ONe deflagration simulations is, rather
unsurprisingly, very similar to the nucleosynthesis in the high density CO deflagration
simulations by \citet{Woosley1997a}. The overabundances (mass fractions relative to solar) of
the stable nuclei after the ejecta has been allowed to decay for $10^{16}$~s are shown in
Figure~\ref{fig:G14a_n100_nkk}. Also shown for reference in Figure~\ref{fig:G14a_n100_nkk} is
model N100DDT from \citet{Seitenzahl2013a}, that follows a delayed detonation in a
Chandrasekhar-mass CO white dwarf star. It is noteworthy that these are of course mass
fractions, and the total ejecta of N100DDT is roughly 50\% more massive than G14a. The four
main distinguishing features of the ONe deflagration compared with N100DDT are:
(1) substantial deficit of C;
(2) ejection of large quantities of O and Ne from the progenitor;
(3) presence of a very large (relative to solar) quantity of trans-iron elements
from Zn to Rb, and
(4) significant overproduction of \Ca{48}, \Ti{50} and \Cr{54}, again relative
to solar.
Observations 1 and 2 are perhaps obvious, and the remaining points have already been identified
and published in the context of high density CO deflagrations by \citet{Woosley1997a}, and so
in this sense what we present here could be conveyed as not particularly novel. However, there
are two aspects of our work that build on \citet{Woosley1997a}: \emph{(a)} the existence of ONe
WDs with such extreme central densities is supported by stellar evolution theory (whereas there
is no clear plausible formation channel for CO WDs with densities as high as
$8\times10^9$~g~cm$^{-3}$), and \emph{(b)} weak reaction rates for the \emph{fp}- and
\emph{fpg}-shell neutron-rich iron-group isotopes are now available.

\citeauthor{Woosley1997a} concluded that given there was no other compelling
site for producing \Ca{48}, exotic high density CO deflagrations must
occasionally occur, at about 2\% of the ``normal'' type~Ia SN rate. However, it
is not completely clear how such a high density CO white dwarf could be formed
without burning C into O and Ne and transforming into an ONe white dwarf.  One
opportunity would be in the merger of two CO white dwarf stars, but those are
expected to either explode during the merging process
\citep[e.g.][]{Pakmor2012a} or also transform into an ONe white dwarf and then a
Si white dwarf, or burning proceeds through to Fe-group elements and the core
collapses into a neutron star \citep[see,
e.g.,][]{Saio1985a,Nomoto1985a,Schwab2016a}.  Our paper proposes a possible
solution to this conundrum in which massive CO white dwarfs well above the
critical mass for carbon ignition and ONe white dwarf formation do not have to
exist. We, of course, are plagued by other, different questions and
uncertainties such as whether or not the conditions of ONe cores at ignition are
favourable for a partial thermonuclear explosion.  If they indeed all collapse,
neither do we have a solution. It should be mentioned, however, that since
\citeauthor{Woosley1997a}'s work in \citeyear{Woosley1997a}, \citet{Wanajo2013b}
found that \Ca{48} can also be produced in ECSNe in the case where they collapse
into a neutron star. The yields from the cECSN simulations from
	\citet{Wanajo2013b,Wanajo2013a} are given in the last column of
	Table~\ref{table:1}, for comparison with our simulation results for
	tECSNe (G14 and J07). Per event, our tECSN simulations produce about
	$2\times10^{-3}~\msun$ of \Ca{48} (ejecta mass is approximately 1~\msun)
	while the cECSN simulations by \citet{Wanajo2013b,Wanajo2013a} produce
	about $2\times10^{-5}~\msun$ -- 2 orders of magnitude less.

\begin{figure*}
	\centering
	\includegraphics[width=\textwidth]{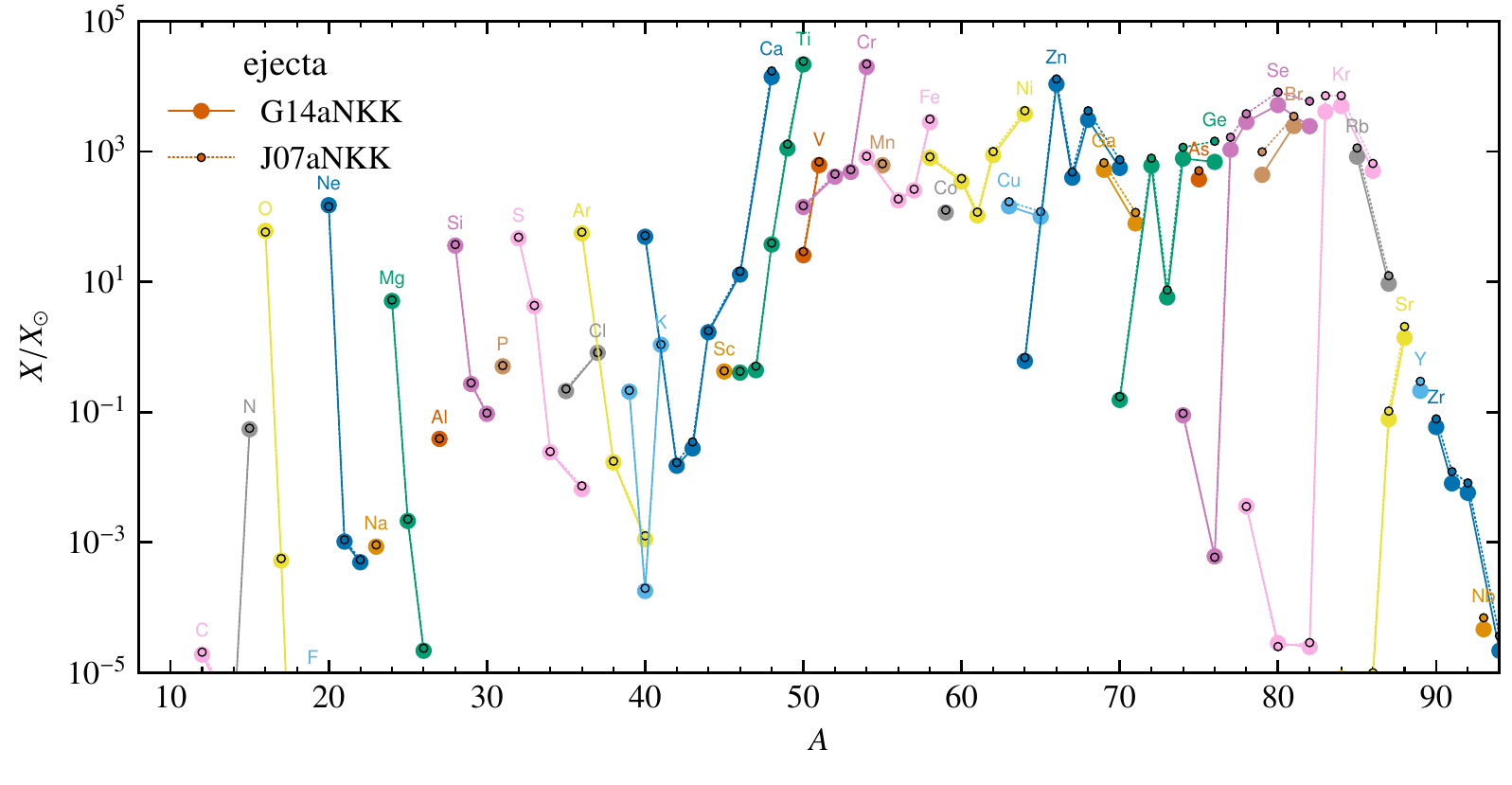}
	\caption{
		Same as Figures~\ref{fig:G14a_n100_nkk} and~\ref{fig:NKK_yield_impact} but
		comparing two simulations with different initial central densities. G14a is the
		$512^3$ simulation from \citet{Jones2016a} with initial central density
		$\log_{10}(\rho/\mathrm{g~cm}^{-3})=9.9$ and J07a is a $576^3$ resolution
		version of simulation J01 from \citet{Jones2016a}, which had an initial central
		density of $\log_{10}(\rho/\mathrm{g~cm}^{-3})=9.95$. The higher density
		simulation exhibits an ejecta that is moderately enhanced in the neutron-rich
		isotopes and the trans-iron elements (see also Table~\ref{table:wdmr-sims} and
		Table~\ref{table:1}).
	}
	\label{fig:density_yield_impact}
\end{figure*}

\begin{table*}
	\caption{Mass fractions of stable isotopes or elements of interest from the solar
	distribution \citep{Asplund2009}, the W7 type~Ia SN model \citep{Nomoto1984b}, the
	N100DDT type~Ia SN model \citep{Seitenzahl2013a}, and the G14a ONe deflagration
	simulation \citep{Jones2016a}, with and without weak reaction rates by
	\citet{Nabi2004a} included in the nucleosynthesis post-processing. The higher-density model J07 is
	also included (see Table~\ref{table:wdmr-sims}). The last column is the cECSN yields
		from \citet{Wanajo2013b,Wanajo2013a}, where the total ejecta
	mass was $1.14\times10^{-2}~\msun$. The simulation
	results are all decayed yields, i.e.~they are for the ejecta only. The N100DDT model
	was decayed to 2~Gyr and the G14 and J07 models to $10^{16}$~s (enough time for all
	radioactive nuclides produced to decay -- see half lives in
	Table~\ref{table:radioactivities}. }
	\label{table:1}
	\centering
	\begin{tabular}{c c c c c c c c}
		\toprule\toprule
		isotope/element & $\odot$ & W7 & N100DDT & G14a & G14aNKK & J07aNKK & W13 \\
		\midrule
		Zn       & 1.85e-06 &  2.06e-05 &  2.22e-06 &  4.83e-03 &  6.80e-03 & 8.33e-03 & 9.97e-02\\
		Se       & 1.34e-07 &  0.00e+00 &  0.00e+00 &  4.51e-05 &  4.80e-04 & 7.60e-04 & 6.87e-03\\
		Kr       & 1.16e-07 &  0.00e+00 &  0.00e+00 &  4.41e-05 &  3.95e-04 & 5.86e-04 & 1.04e-02\\
		\Ca{48}  & 1.53e-07 &  1.90e-09 &  5.60e-15 &  1.32e-03 &  2.13e-03 & 2.64e-03 & 2.04e-03\\
		\Ti{50}  & 1.79e-07 &  7.82e-05 &  1.86e-07 &  4.46e-03 &  3.91e-03 & 4.37e-03 & 1.65e-04\\
		\Cr{54}  & 4.33e-07 &  6.75e-04 &  7.92e-06 &  9.48e-03 &  8.64e-03 & 9.56e-03 & 3.97e-04\\
		\bottomrule\bottomrule
	\end{tabular}
\end{table*}

\begin{table}
	\caption{Ejected masses of radioactive isotopes in the 
		N100DDT type~Ia SN model \citep{Seitenzahl2013a}, and the G14aNKK ONe
		deflagration simulation \citep{Jones2016a} ejecta at 100~s after ignition.
		Isotopes with half lives less than $10^4$~s have been omitted.
	}
	\label{table:radioactivities}
	\centering
	\begin{tabular}{cccc}
		\toprule\toprule
		isotope & N100DDT & G14aNKK & half life / s \\
		\midrule
		$^{56}$Ni &     6.04e-01 &     1.87e-01 &     5.25e+05 \\ 
		$^{57}$Ni &     1.79e-02 &     6.00e-03 &     1.28e+05 \\ 
		$^{66}$Ni &              &     5.37e-03 &     1.97e+05 \\ 
		$^{55}$Co &     1.14e-02 &     4.45e-03 &     6.31e+04 \\ 
		$^{60}$Fe &     4.20e-10 &     2.81e-03 &     8.27e+13 \\ 
		$^{52}$Fe &     7.93e-03 &     2.42e-03 &     2.98e+04 \\ 
		$^{55}$Fe &     1.86e-03 &     1.48e-03 &     8.66e+07 \\ 
		$^{57}$Co &     8.70e-04 &     6.85e-04 &     2.35e+07 \\ 
		$^{59}$Ni &     3.93e-04 &     2.65e-04 &     2.40e+12 \\ 
		$^{53}$Mn &     2.35e-04 &     1.72e-04 &     1.18e+14 \\ 
		$^{62}$Zn &     3.22e-04 &     1.36e-04 &     3.31e+04 \\ 
		$^{56}$Co &     1.18e-04 &     8.78e-05 &     6.67e+06 \\ 
		$^{48}$Cr &     3.14e-04 &     8.63e-05 &     7.76e+04 \\ 
		$^{59}$Fe &     2.72e-09 &     5.26e-05 &     3.84e+06 \\ 
		$^{72}$Zn &              &     3.98e-05 &     1.67e+05 \\ 
		$^{63}$Ni &     1.76e-08 &     2.59e-05 &     3.19e+09 \\ 
		$^{67}$Cu &              &     2.00e-05 &     2.23e+05 \\ 
		$^{77}$Ge &              &     9.99e-06 &     4.04e+04 \\ 
		$^{54}$Mn &     3.03e-06 &     7.57e-06 &     2.70e+07 \\ 
		$^{58}$Co &     4.35e-06 &     4.18e-06 &     6.12e+06 \\ 
		$^{51}$Cr &     9.29e-06 &     3.59e-06 &     2.39e+06 \\ 
		$^{44}$Ti &     9.98e-06 &     2.50e-06 &     1.87e+09 \\ 
		$^{52}$Mn &     5.18e-06 &     2.12e-06 &     4.83e+05 \\ 
		$^{37}$Ar &     3.43e-05 &     1.70e-06 &     3.02e+06 \\ 
		$^{60}$Co &     2.03e-08 &     1.10e-06 &     1.66e+08 \\ 
		$^{61}$Cu &              &     9.58e-07 &     1.20e+04 \\ 
		$^{85}$Kr &              &     4.03e-07 &     3.39e+08 \\ 
		$^{41}$Ca &     6.07e-06 &     2.36e-07 &     3.14e+12 \\ 
		$^{73}$Ga &              &     7.06e-08 &     1.75e+04 \\ 
		$^{47}$Ca &              &     6.69e-08 &     3.92e+05 \\ 
		$^{49}$V  &     3.57e-07 &     5.01e-08 &     2.85e+07 \\ 
		$^{88}$Kr &              &     4.98e-08 &     1.02e+04 \\ 
		$^{48}$V  &     9.12e-08 &     4.41e-08 &     1.38e+06 \\ 
		$^{77}$As &              &     4.35e-08 &     1.40e+05 \\ 
		$^{22}$Na &     4.27e-09 &     4.10e-08 &     8.21e+07 \\ 
		$^{26}$Al &     5.68e-07 &     2.42e-08 &     2.26e+13 \\ 
		$^{79}$Se &              &     2.02e-08 &     1.03e+13 \\ 
		$^{45}$Ti &              &     1.97e-08 &     1.11e+04 \\ 
		$^{64}$Cu &              &     3.86e-09 &     4.57e+04 \\ 
		$^{48}$Sc &              &     3.56e-09 &     1.57e+05 \\ 
		$^{43}$Sc &              &     2.60e-09 &     1.40e+04 \\ 
		$^{47}$Sc &              &     2.36e-09 &     2.89e+05 \\ 
		$^{90}$Sr &              &     7.32e-10 &     9.09e+08 \\ 
		$^{65}$Zn &     7.35e-10 &     3.65e-10 &     2.11e+07 \\ 
		$^{66}$Ga &              &     3.39e-10 &     3.42e+04 \\ 
		$^{68}$Ge &     6.33e-10 &     1.91e-10 &     2.34e+07 \\ 
		$^{89}$Sr &              &     1.06e-10 &     4.37e+06 \\ 
		$^{36}$Cl &     7.77e-07 &     1.03e-11 &     9.51e+12 \\ 
		$^{33}$P  &     3.76e-07 &     9.77e-12 &     2.19e+06 \\ 
		$^{32}$Si &     9.47e-09 &     7.32e-12 &     4.83e+09 \\ 
		$^{35}$S  &     5.39e-07 &     3.04e-12 &     7.55e+06 \\ 
		$^{32}$P  &     4.96e-07 &     1.97e-12 &     1.23e+06 \\ 
		$^{40}$K  &     5.81e-08 &     9.99e-13 &     3.94e+16 \\ 
		$^{39}$Ar &     1.29e-08 &     3.17e-13 &     8.49e+09 \\ 
		$^{14}$C  &     2.47e-06 &     6.49e-17 &     1.80e+11 \\ 

		\bottomrule\bottomrule
	\end{tabular}
\end{table}

We were fortunate enough to have access to the QRPA calculations by \citet{Nabi2004a} for the
\emph{fp} and \emph{fpg} shell nuclei on the neutron-rich side of the valley of stability,
which were not available in \citeyear{Woosley1997a} when \citeauthor{Woosley1997a} conducted
his study of deflagrations in high-density CO white dwarfs. It is worth also mentioning that
many of the reaction rates that we have used originate from more recent measurements or
calculations than those used by \citet{Woosley1997a}. Of particular note are the weak reaction
rates for the \emph{pf} shell nuclei by \citet{Langanke2000}. \citet{Woosley1997a} commented that at
some point when weak reaction rates for the \emph{fp} and \emph{fpg} shell nuclei became available, it
would be of some interest to study how their inclusion could change the nucleosynthesis yields
from deflagrations in high-density CO white dwarfs. We have indeed done this, but for high
density ONe white dwarfs.  We expect that the outcome is probably very similar whether the fuel
is CO or ONe, because (a) the binding energy (relative to free nucleons) of $^{12}$C is similar
to $^{20}$Ne, i.e.  $7.41\times10^{18}$~erg~g$^{-1}$ for $^{12}$C and
$7.75\times10^{18}$~erg~g$^{-1}$ for $^{20}$Ne (numbers are relative to free nucleons) and (b)
because for both CO and ONe white dwarfs, at least 50\% of the mass is ususally $^{16}$O. In
fact, the binding energy of $^{16}$O is $7.70\times10^{18}$~erg~g$^{-1}$ (99\% that of
$^{20}$Ne), meaning that an ONe white dwarf is very similar indeed to a CO white dwarf with a
low C/O ratio\footnote{The C/O ratio resulting from He burning is very sensitive to the
	$^{12}$C$(\alpha,\gamma)^{16}$O reaction rate. See \citet{deBoer2017} for a recent
thorough review of this reaction from a nuclear physics perspective}.

The impact of including the NKK04 reaction rates in the post-processing nucleosynthesis
simulation of model G14a is shown in Figure~\ref{fig:NKK_yield_impact}. As we have shown in
Figure~\ref{fig:ye_with_NKK}, including the NKK04 rates results in faster deleptonization at
high densities than when they are omitted. We also showed in Figure~\ref{fig:NSE} that with
decreasing \ye~(for fixed $T,\rho$), the NSE distribution solution favours not only more
neutron-rich nuclei, but nuclei with higher atomic weight, than at higher \ye. This effect is
evident in Figure~\ref{fig:NKK_yield_impact} in the extra production of the trans-iron elements
between Zn and Sr.  Although the changes may not look like much in
Figure~\ref{fig:NKK_yield_impact}, because of the logarithmic scale, the enhancement of the
elemental abundances of both Se and Kr are about 1~dex when the NKK04 rates are included (see
Table~\ref{table:1}). The \Ca{48} yield increases by 61\%, the Zn yield increases by 41\% and
the yields of \Ti{50} and \Cr{54} decrease by 12\% and 8.9\%, respectively. The final mass
fractions of \Ca{48}, \Ti{50} and \Cr{54} for the tracer particles in the G14 simulation
experiencing the most extreme conditions are shown as a function of the peak temperature and
peak density in Figure~\ref{fig:tps}.

Also shown in Table~\ref{table:1} are the mass fractions of these isotopes and
elements in the ejecta of simulation J07a (final column). This simulation is a
$576^3$ version of the simulation J01 from \citet{Jones2016a}, which had an
initial central density of $\log_{10}(\rho/\mathrm{g~cm}^{-3})=9.95$, compared
to $9.9$ for G14a. One can see from the comparison in
Figure~\ref{fig:density_yield_impact} that the impact of the higher initial
density is a moderate enhancement of the trans-iron elements and the
neutron-rich isotopes \Ca{48}, \Ti{50} and \Cr{54}. Otherwise, the abundance
distribution in the two models looks very similar. This implies that the central
density of the ONe core when the deflagration wave is ignited by $^{20}$Ne
electron captures is a secondary effect in determining the distribution of the
composition in the ejecta. We have not yet fully tested the impact of varying
the ignition geometry (the position and shape of the initial flame kernels) on
the ejecta composition, but we estimate that this will likely not have much of
an effect.

The ejected masses of several radioactive isotopes produced in the ejecta of simulation G14 are
given in Table~\ref{table:radioactivities}, in descending order of ejected mass, together with
their half-lives. The respective numbers from the DDT simulation N100DDT from
\citet{Seitenzahl2013a} are also given, for comparison.  One of the more interesting signatures
of the composition of the G14 ejecta is the exceptionally large ratio of the two long-lived
radionuclides $^{60}\mathrm{Fe}/^{26}\mathrm{Al}$. In the G14a, the ratio of their mole
fractions is $Y(^{60}\mathrm{Fe})/Y(^{26}\mathrm{Al}) = 4.94\times10^4$. If one includes the
$\sim10^{-5}~M_\odot$ of $^{26}$Al from the envelope of the SAGB star \citep{Siess2008a}, this
becomes $Y(^{60}\mathrm{Fe})/Y(^{26}\mathrm{Al}) \approx 130$.  The INTEGRAL/SPI mission has
measured the line flux ratio $F(^{60}\mathrm{Fe})/F(^{26}\mathrm{Al})$ in the diffuse
interstellar medium to be 0.17 \citep[][but see \citealp{Wang2007a} for a discussion of several
similar measurements]{Bouchet2011a} -- three to five orders of magnitude lower. The predominant
source of both $^{60}$Fe and $^{26}$Al is thought to be massive stars and their FeCCSNe. The
same ratio from FeCCSNe is typically between 0.1 and 1 \citep[see, e.g.][]{Timmes1995a}, making
the ratio in our simulations something quite unique.  There are two main reasons for this.
First of all, $^{26}$Al is produced in the H-burning, Ne-burning and O-burning shells in
massive stars, with an additional contribution from shock and neutrino nucleosynthesis in the
Ne and O shells.  Of course, in our ONe deflagration simulations we are considering only the
ONe white dwarf and therefore there is no $^{26}$Al from H burning, although for ECSNe from
single stars, there will be a contribution from the H envelope. To continue the comparison with
massive stars, the Ne and O burning in a tECSN predominantly reaches NSE at the deflagration
front. As one can see from Figure~\ref{fig:NSE}, sd-shell nuclei such as $^{26}$Al (or
$^{25}$Mg, from which $^{26}$Al can be created via $(p,\gamma)$), are not terribly abundant in
the NSE compositions we encounter, particularly below $\ye=0.5$. Another contrasting feature
between massive stars/FeCCSNe and tECSNe is the mechanism of $^{60}$Fe production. $^{60}$Fe is
produced during the $s$ process in core He burning and C shell burning in massive stars and
proceeds by neutron capture on $^{59}$Fe, where the neutrons are released by the
$^{22}$Ne$(\alpha,n)^{25}$Mg reaction \citep[see, e.g.][]{Timmes1995a,Limongi2006a,Tur2010a}.
The same reaction sequence takes place during the FeCCSN as the shock passes through the C
shell and the He shell, only on much shorter time-scales and much higher neutron densities than
in the $s$ process. In the tECSNe, $^{60}$Fe is produced in the NSE state behind the
deflagration front. This is most effective when $\ye\approx26/60\approx0.43$, and such a low
\ye~is obtained in ONe deflagrations but not in normal Type Ia supernovae.  The
implications for the $F(^{60}\mathrm{Fe})/F(^{26}\mathrm{Al})$ ratio in the interstellar medium
(ISM) could also be used as a constraint for the rate of occurence of tECSNe, however we
believe that the current uncertainties in massive star yields for these two radionuclides
(Wolf-Rayet mass loss rates and the currently unmeasured $^{59}$Fe$(n,\gamma)^{60}$Fe cross
section) prevent this constaint from being particularly meaningful at present. Indeed,
current massive star models generally produce ratios that are too large to explain the INTEGRAL
measurement.

\begin{figure}
	\centering
	\includegraphics[width=\linewidth]{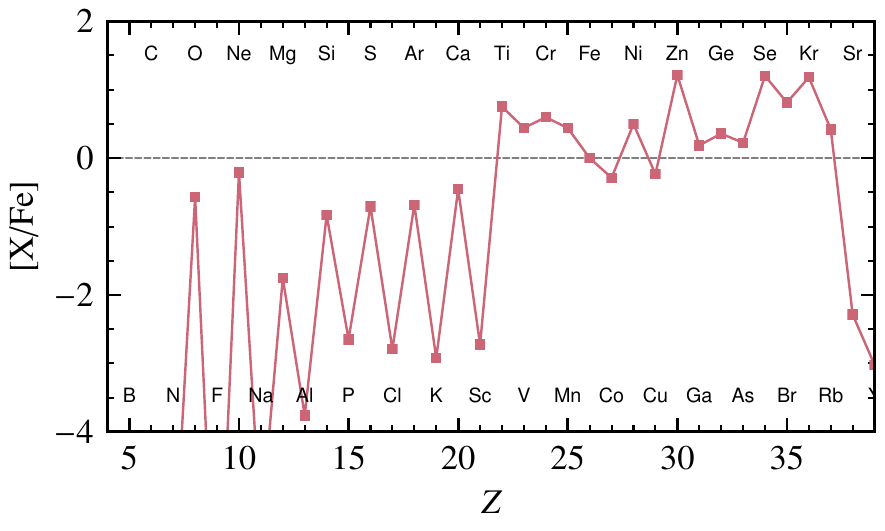} \\
	\caption{
		Elemental yields from simulation G14aNKK relative to Fe and normalized to the
		solar composition.
	}
	\label{fig:G14a-el}
\end{figure}

The elemental yields for the simulation G14aNKK are presented in Figure~\ref{fig:G14a-el}. The
large production of Zn compared to Fe in the simulations is a feature in common with hypernovae
(HNe), which are currently the most favourable scenario to explain the high [Zn/Fe] observed in
the oldest stars in the Milky Way \citep[e.g.,][and references
therein]{Kobayashi2011a,Nomoto2013a}. Based on Figure~\ref{fig:G14a-el}, there might be the
possibility that ECSNe could be an additional or even dominant source of Zn. This will of
course need to be explored further and in more detail. The large production of trans-Fe
elements relative to Fe, particularly Se and Kr, may limit the amount of Zn that could
come from tECSNe in the early Galaxy.  The tECSN simulations also show a strong production of
Ti and Mn. The ratios [Ti/Fe] and [Mn/Fe] are currently not well reproduced in GCE simulations
at low metallicities. Theoretical GCE simulations considering FeCCSNe and HNe contribution only
tend to underestimate Ti and Mn compared to the observations of the majority of metal
poor-stars \citep[e.g.,][]{kobayashi:11,Sneden2016a}. The role of tECSNe in contributing to
these elements in a chemical evolution context is therefore also something we would like to
explore in the future.

\section{Binary population synthesis simulations}
\label{sec:popsynth}

In reality many of the progenitor stars of ECSNe will exist in close,
interacting binary systems, which must be taken into account when predicting the
frequency of their occurrence.  We used the binary evolution population synthesis
code {\sc StarTrack} \citep[e.g.][]{belczynski2002a,belczynski2008a} to
calculate birthrates of ECSNe arising from single and binary stars assuming
field-like (no dynamics) evolution.\footnote{Neglecting ECSNe formed in dense
	stellar environments like globular clusters is a valid assumption since
	only a small fraction of stellar mass exists in globular clusters.} We
	also included rates of accretion-induced collapse (AIC) of accreting
	white dwarfs as presented in \citet{ruiter2018a}, where an oxygen-neon
	white dwarf approaches the Chandrasekhar mass via accretion (RLOF or
	wind-accretion) from a stellar companion.

	Following \citet{hurley2000a}, an ECSN is identified based on a star's
	He core mass at the base of the asymptotic giant branch ($M_{\rm
	He,BAGB}$). We used the calculations of
	\citet{EldridgeTout2004b,EldridgeTout2004a} to allow for ECSNe for
	$M_{\rm He,BAGB}=1.83-2.25~\msun$. This corresponds to Zero Age Main
	Sequence star mass range $M_{\rm ZAMS}=7.6-8.3~\msun$ for solar-like
	metallicty ($Z=0.02$) for single stars.  The evolution of single stars
	was performed with analytic fits to detailed stellar models
	\citep{hurley2000a}, with an updated wind mass loss prescription
	\citep{belczynski2010a}. In binary evolution we used the same range of
	the He core mass to decide when we encounter an ECSN. However, we note that
	during binary evolution mass gain and mass loss during Roche lobe
	oveflow may affect the initial ZAMS mass range for which an ECSN is
	encounetred, generally making it broader than for single stars. The
	details of the binary evolutionary prescriptions are described in
	\citet{belczynski2008a}.

For this paper we employed the same prescription for common envelope evolution as
described in \citet[][the `new CE' model, where the binding energy parameter $\lambda$ is
dependent on the evolutionary stage of the donor]{ruiter2018a}, and all stars
were evolved with
an initial near-solar metallicity ($Z=0.02$). However, the simulations discussed in this paper
differ in their initial orbital parameter distributions. While we assumed a three-component IMF
for single stars and primary stars in binaries \citep[see][]{ruiter2009a}, the secondary stars
in binaries, rather than being drawn from a flat mass ratio distribution, were drawn from a
distribution based on \citet{Sana2012a}. While a flat mass ratio distribution has been the
general standard widely adopted in population synthesis studies of low- and intermediate-mass
stars, with new observational analyses it is becoming clear that some of the standard choices
for theoretically-adopted orbital parameters require some re-evaluation \citep[see
e.g.][]{moe2017a}. Since we want to compare our ECSN rates with rates of core-collapse SNe, we
adopted the \citet{Sana2012a} probability distribution functions for our simulations since these
distributions were found to be very important for massive stars. Following \citet{Sana2012a},
we adopted initial period and eccentricity power-law distributions accordingly.
We assumed a conservative binary fraction of 50\% to calibrate our numbers, meaning we assume
that for every single star produced, a binary is produced.

We present ECSN birthrates in Table~\ref{table:startrack} normalized by total mass formed in
stars (assuming a mass range of $0.08 - 150$ \msun), and also relative to the total number of
core collapse supernovae \citep[see][for treatment of core-collapse supernovae and
ECSNe]{chruslinska2018}. The total rate for all AICs and ECSNe from single stars and stars in
binary systems is 3.31~\% of the FeCCCN rate, with the majority being ECSNe from stars in
binary systems (occurring at 2.8~\% of the FeCCSN rate). In the following section we will
estimate an upper limit for this rate from the results of our tECSN nucleosynthesis simulations
using the solar abundance distribution and show that the upper limit is in relatively good
agreement with the population synthesis results.

\begin{table}
	\caption{Relative total number of events from {\sc StarTrack} that occur per
	simulation of 5.12 million ZAMS binaries and 5.12 million ZAMS single stars.
	We show birthrates of accretion-induced collapse ONe WDs to NSs (AIC), ECSNe
	from binaries and ECSNe from single stars. The rates are presented per total
	stellar mass formed in stars (rate \msun $^{-1}$) and relative to the total
	core collapse supernova rate from the same simulated population (percentages
	given in braces). A 50\% binary fraction is assumed (see text).
    }
	\label{table:startrack}
	\centering
	\begin{tabular*}{\linewidth}{l @{\extracolsep{\fill}} c c c}
                \toprule\toprule
		Event       & Rate \msun$^{-1}$ & \multicolumn{2}{c}{Rate rel to CCSN} \\
		\midrule
		\multicolumn{4}{c}{} \\
		AIC         & 1.9e-5                  & 4e-3 & (0.36 \%) \\
		ECSN binary & 1.4e-4                  & 3e-2 & (2.8 \%) \\
		ECSN single & 1.5e-5                  & 2e-3 & (0.15 \%) \\
		\midrule
		total       & 1.7e-4                  & 3.6e-2 & (3.31 \%) \\
		\bottomrule \bottomrule
	\end{tabular*}
\end{table}

\begin{table*}
	\caption{Mass fractions of stable isotopes or elements of interest relative to $^{16}$O
		from the solar distribution \citep{Asplund2009}, the simulation G14a from
		\citet{Jones2016a} post-processed including weak reaction rates from
		\citet{Nabi2004a}, and the IMF-weighted average per-explosion FeCCSN ejecta
		from \citet[][at $Z=0$ and $Z=0.001$ and $Z=0.004$]{Nomoto2006}
		together with the maximum allowed number of ECSNe as a fraction of the number
		of FeCCSNe for each set of assumed FeCCSN yields.
	}
	\label{table:2}
	\centering
	\begin{tabular}{c c c c c c c c c}
		\toprule\toprule
		isotope/element & $\left(\ddfrac{X_i}{X(^{16}\mathrm{O})}\right)_\odot$ &
		G14aNKK &
		\multicolumn{2}{c}{N06 ($Z=0$)} &
		\multicolumn{2}{c}{N06 ($Z=0.001$)} &
		\multicolumn{2}{c}{N06 ($Z=0.004$)} \\
		&&&
		$X_i/X(^{16}\mathrm{O})$ &
		$\left(\ddfrac{N_\mathrm{ECSN}}{N_\mathrm{CCSN}}\right)_\mathrm{max}$ &
		$X_i/X(^{16}\mathrm{O})$ &
		$\left(\ddfrac{N_\mathrm{ECSN}}{N_\mathrm{CCSN}}\right)_\mathrm{max}$ &
		$X_i/X(^{16}\mathrm{O})$ &
		$\left(\ddfrac{N_\mathrm{ECSN}}{N_\mathrm{CCSN}}\right)_\mathrm{max}$ \\
		\midrule
		\Ca{48}  &  2.52e-05 & 5.79e-03 & 6.87e-13 & 2.63e-02 & 9.93e-08 & 2.72e-02 & 3.04e-07 & 2.34e-02 \\
		\Ti{50}  &  2.94e-05 & 1.06e-02 & 9.12e-13 & 1.69e-02 & 2.98e-07 & 1.74e-02 & 1.13e-06 & 1.47e-02 \\
		\Cr{54}  &  7.12e-05 & 2.35e-02 & 4.92e-09 & 1.85e-02 & 8.02e-07 & 1.90e-02 & 2.56e-06 & 1.61e-02 \\
		\Zn{66}  &  8.55e-05 & 1.54e-02 & 2.07e-07 & 3.33e-02 & 7.33e-06 & 3.18e-02 & 2.68e-05 & 2.10e-02 \\
		\Zn{67}  &  1.27e-05 & 8.37e-05 & 4.82e-09 & 5.26e-01 & 9.38e-07 & 5.17e-01 & 4.40e-06 & 3.96e-01 \\
		\Zn{68}  &  5.92e-05 & 3.02e-03 & 8.56e-09 & 1.10e-01 & 6.27e-06 & 1.03e-01 & 2.71e-05 & 5.69e-02 \\
		\Zn{70}  &  2.01e-06 & 1.87e-05 & 5.01e-15 & 4.27e-01 & 5.87e-08 & 4.29e-01 & 1.44e-07 & 3.84e-01 \\
		\bottomrule\bottomrule
	\end{tabular}
\end{table*}

\section{Occurrence rate constraints from abundance measurements}

In this section we estimate the maximum rate at which tECSNe could occur without overproducing
\Ca{48}, \Ti{50}, \Cr{54} and \Zn{66} (since these have the largest abundances relative to
solar in the ejecta; see Figure~\ref{fig:G14a_n100_nkk}). This is done by combining our tECSN
yields presented in Section~\ref{sec:yields} with Salpeter IMF-weighted FeCCSN yields from
\citet{Nomoto2006} and comparing the resulting composition to the solar abundance distribution.
The maximum tECSN rate is found to be consistent with the ECSN rates from population synthesis
simulations from this work (Section~\ref{sec:popsynth}) and from stellar evolution models by
\citet{Poelarends2007,Poelarends2008} and \citet{Doherty2015,Doherty2017a}.

Because ECSNe are typically thought to collapse into NSs, their estimated rate is usually given
as a fraction of all core-collapse events, where core-collapse events includes ECSNe and
FeCCSNe. Although in this work we are considering the case for which ECSNe do not result
in core-collapse, we still stick to convention to make a comparison with statistics from other
studies. We define $f$ to be the number of ECSNe as a fraction of the total number of (ECSN+FeCCSN) events,
\begin{equation}
	f = \dfrac{N_\mathrm{EC}}{N_\mathrm{EC}+N_\mathrm{CC}}.
\end{equation}
Since we expect that the number of ECSNe is much lower than the number of CCSNe,
$N_\mathrm{EC} \ll N_\mathrm{CC}$ (see, e.g. Table~\ref{table:startrack}), to
fairly good approximation
\begin{equation}
	f\approx\dfrac{N_\mathrm{EC}}{N_\mathrm{CC}}.
\end{equation}
Therefore, we will discuss the fraction $f$ as being the number of ECSNe relative to the number
of CCSNe, or the rate of ECSNe relative to the CCSN rate. We feel that clarifying this point
will make the discussion easier to follow and will make the comparison with the ECSN rate
predictions from stellar evolution, cECSN nucleosynthesis and population synthesis
syntactically more straightforward.

With this definition of $f$, the following equality should be true for two isotopes $i$ and $j$
made only in ECSNe and FeCCSNe:
\begin{equation}
	\label{eq:gce}
	\left(\ddfrac{M^i}{M^j}\right)_\odot =
	\ddfrac{(1-f)\bar{M}^i_\mathrm{CC} + fM^i_\mathrm{EC}}
	{(1-f)\bar{M}^j_\mathrm{CC} + fM^j_\mathrm{EC}}.
\end{equation}
If an isotope is also partially produced in a site other than ECSNe or FeCCSNe, then the solar
ratio (LHS of Equation~\ref{eq:gce}) is an upper limit, and the RHS should remain below it. In
either case, it is important that the ratio does not exceed the solar ratio. We
	consider the isotopes \Ca{48}, \Ti{50} and \Cr{54}, \Zn{66}, \Zn{67}, \Zn{68} and
	\Zn{70} and their abundances relative to the abundance of $^{16}$O.  We are therefore
	assuming that these isotopes are produced in, and only in, FeCCSNe and tECSNe,
	implying a negligible contribution to the solar inventory from ``normal'' Type Ia SNe
	or AGB stars. This assumption is pretty sound for SNe Ia for all the isotopes
	considered here. For AGB stars, this is also a sound assumption for \Ca{48}, \Ti{50}
	and \Cr{54}. For the Zn isotopes the assumption that the solar inventory of Zn comes
	from FeCCSNe and ECSNe is good to about 10~\% or better. That is, the contribution of
	the main s-process in AGB stars to the solar inventory of Zn is of the order of 10~\%
	or less \citep{Bisterzo2014a}.

It is important
to clarify that we have assumed the same amount and composition of ejecta for all ECSNe (EC),
and that all ECSNe are tECSNe whose yields are given by our nucleosynthesis simulations
(including NKK04 weak rates) of model G14 by \citet{Jones2016a}, and for the FeCCSNe (CC) we
use a single ejecta mass and composition that is the IMF-weighted average
\begin{equation}
	\label{eq:mccsnint}
	\bar{M}^i_\mathrm{CC} =
	\ddfrac{\int_{m_l}^{m_u}\mej^i(\mini)\xi(\mini)~\mathrm{d}\mini}
	{\int_{m_l}^{m_u}\xi(\mini)~\mathrm{d}\mini},
\end{equation}
where $\xi(\mini)$ is the initial mass function $\xi(M)=\xi_0M^{-\alpha}$ with $\alpha = 2.35$
\citep{Salpeter1955} and $\mej^i(\mini)$~is the ejected mass of the isotope or element $i$
collectively in the stellar wind and the FeCCSN of a star with initial mass \mini. $\xi_0$ is a
constant related to the local stellar density. The integral limits are the bounding intial
masses of stars that undergo FeCCSN. That is, $m_l$ is the delimiting mass in-between
super-AGB stars and massive stars that will undergo core-collapse, and $m_u$ is the delimiting
mass in-between massive stars that will undergo core-collapse and massive stars that will
become unstable to the pair creation and become pulsational pair-instability supernovae.

\begin{figure*}
	\centering
	\includegraphics[width=.95\textwidth]{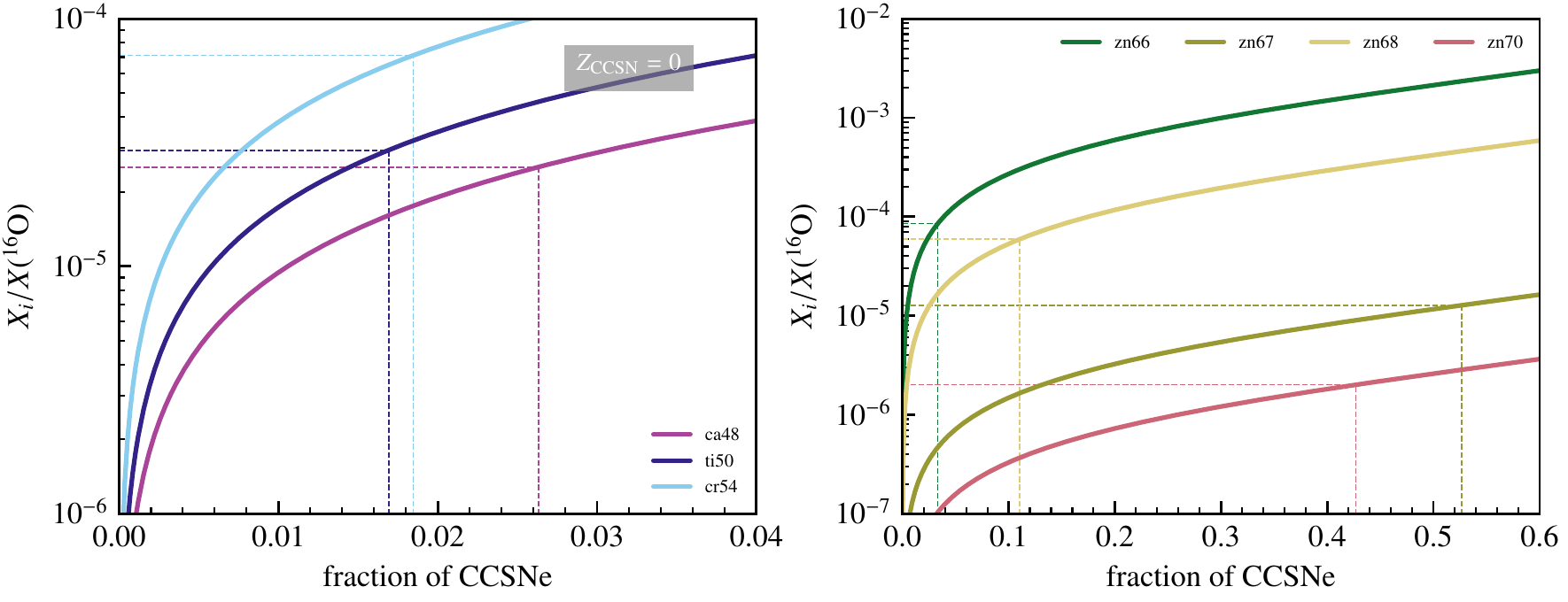} \\
	\includegraphics[width=.95\textwidth]{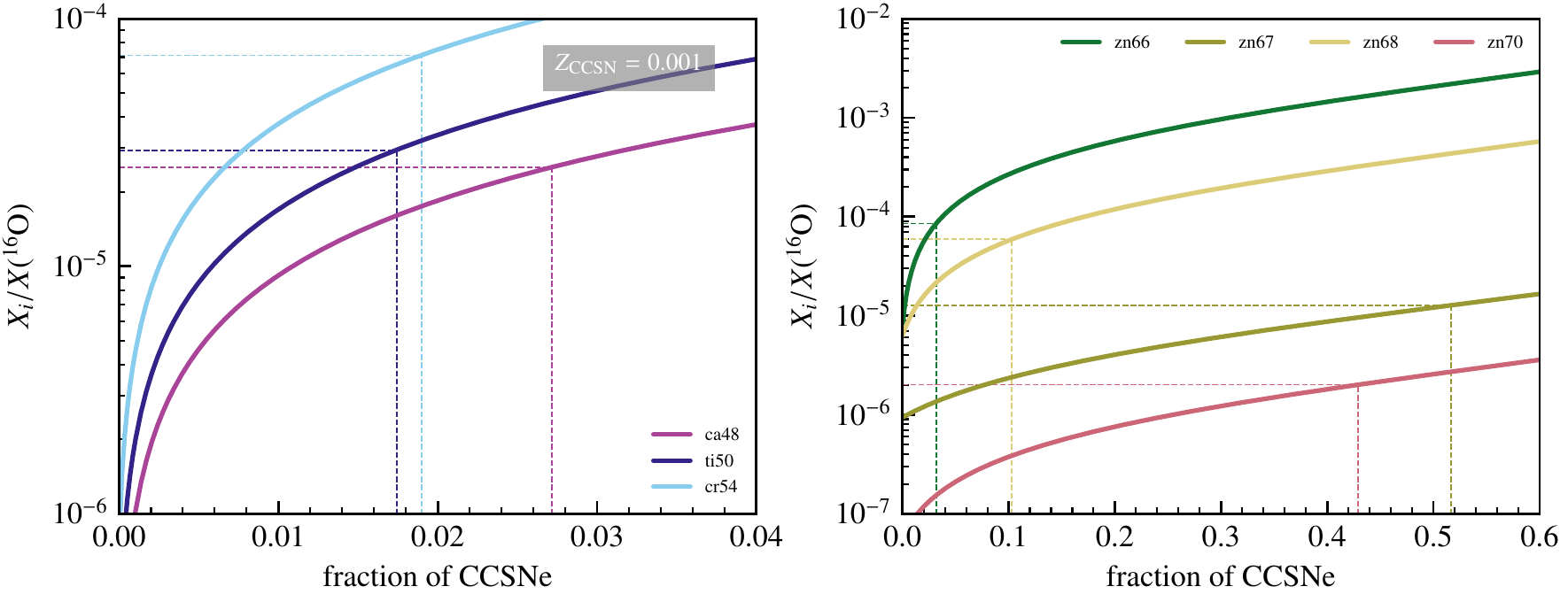} \\
	\includegraphics[width=.95\textwidth]{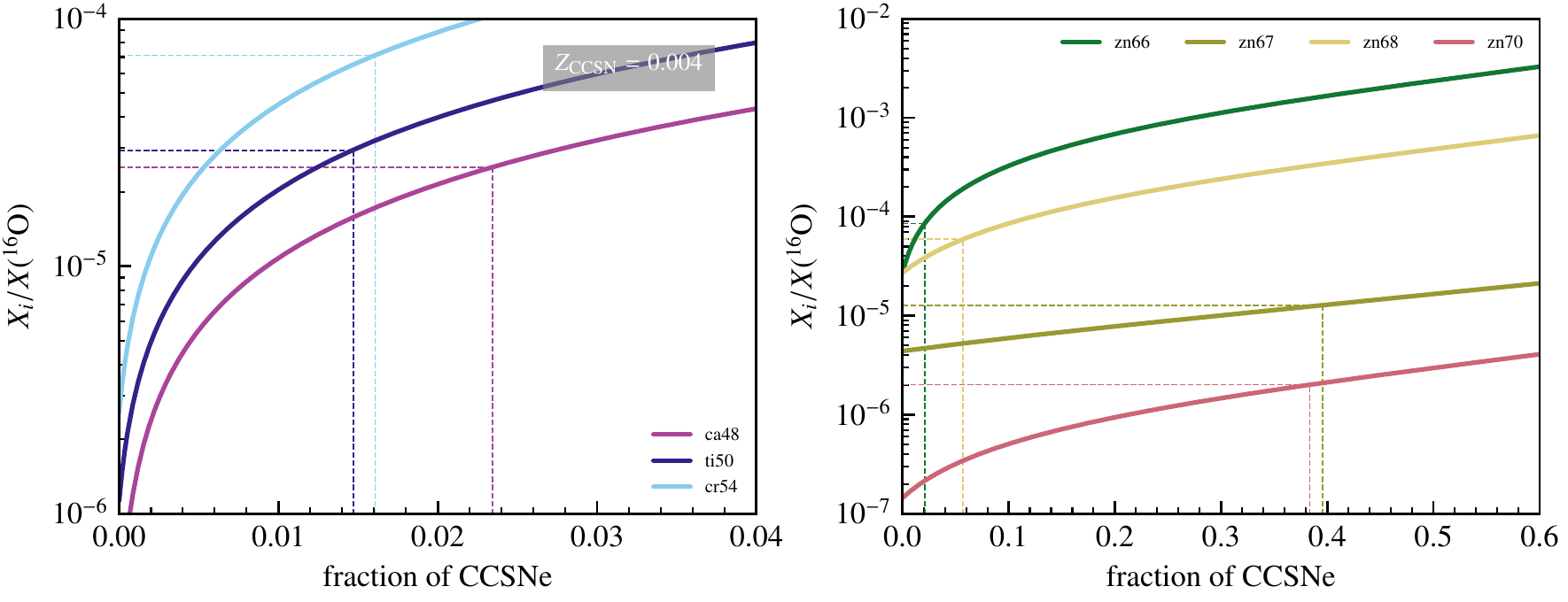}
	\caption{
	Number of tECSNe as a fraction of FeCCSNe, constrained by the solar ratio of key
	isotopes produced in the ejecta to oxygen. The minimum mass for FeCCSN
	was assumed to be
	$m_l=9~\msun$.  The dashed horizontal lines are the solar value for each isotope, which
	we take as an upper limit. The dashed vertical lines therefore indicate the upper limit
	on the number of ECSNe using each constraint from the solar abundances. The CCSN yields
	used were those of \citet{Nomoto2006} at metallicity $Z=0$ (top row), $Z=0.001$ (middle
	row) and $Z=0.004$ (bottom row).
	}
	\label{fig:snfrac}
\end{figure*}

The lower mass limit for stars that explode as FeCCSNe was assumed to be $m_l=9~\msun$ for this
study \citep[we refer the reader to the following several relevant and recent publications
regarding this mass limit:][]{Jones2013,Doherty2015,Doherty2017a,Woosley2015a}. The upper limit
for the initial mass of stars that explode as FeCCSNe was taken to be $m_u=80~\msun$. This is
almost certainly above the initial mass for which black holes are expected to form, but below
the initial mass of stars that are expected to undergo pulsational pair instabilities
\citep[see, e.g.][]{Woosley2017a}. In fact, the yield sets that we have used for FeCCSNe come
from \citet{Nomoto2006} and the most massive star for which yields are provided is 40~\msun.
This means that we have assumed that all stars from 35~\msun~to 80~\msun~have the same yields,
given by the 40~\msun~model.

In order to compute $\bar{M}^i_\mathrm{CC}$ we need the complete yields of stars with initial
masses in the range $m_l \leq \mini \leq m_u$. These are only available as discrete data points
in initial mass space and so we can either interpolate the data in-between the points (e.g.
trapezoidal numerical integration or something more sophisticated) or we can bin the data and
assume that the data points are average values for the bin. This second approach is the most
common practice in galactic chemical evolution, because it prevents the artificial introduction
of new extrema into the data set. However, it is also a less-than-satisfactory practice because
there is likely quite a large variation in ejecta mass and composition as a function of
progenitor mass, particularly at the low-mass end of the FeCCSN progenitor mass range.
Nevertheless, this is only one of the many challenges of chemical evolution. Using the
binned data, Equation~\ref{eq:mccsnint} becomes
\begin{equation}
	\bar{M}^i_\mathrm{CC} = 
	\ddfrac{\sum_{j=1}^{N}\left(\mej^i(\mini^j)\int_{M_{j-1/2}}^{M_{j+1/2}}\xi(\mini)~\mathrm{d}\mini\right)}
	{\int_{m_l}^{m_u}\xi(\mini)~\mathrm{d}\mini},
\end{equation}
where N is the number of mass bins, $M_{j-1/2}$ and $M_{j+1/2}$ are the edges of each mass bin
$j$ and $\mej^i(\mini^j)$ is the ejected mass of isotope $i$ in the wind and FeCCSN of a star
with mass $\mini^j$, which is at the bin centre and is assumed to be the average for the whole
bin.

In Figure~\ref{fig:snfrac} the ratios of the masses of our chosen isotopes to the mass
of $^{16}$O in the mixed ejecta of tECSNe and FeCCSNe for some hypothetical population of stars
(RHS of Equation~\ref{eq:gce}) are plotted against the fraction of (ECSNe + FeCCSNe) that
constitute tECSNe in this hypothetical population ($f$ from Equation~\ref{eq:gce}).  The
horizontal lines demarcate the corresponding solar ratio $X_i/X(^{16}\mathrm{O})$ taken from
\citet{Asplund2009}, which is our upper limit from Equation~\ref{eq:gce}. That is,
values of $f$ for which a ratio exceeds this limit are inconsistent with the chemical
evolution leading to the formation of the Sun for the chosen set of FeCCSN
yields. We have excluded the
yields for FeCCSNe from massive stars with $Z=0.02$ from \citet{Nomoto2006}, even though they
are provided, because they would be inconsistent with the evolution of a population of stars
from whose mixed ejecta the Sun was formed.

The most stringent constraint from the set of isotopes that we have considered comes from
\Ti{50} for any of the three sets of FeCCSN yields we have used. The upper limit for the rate
of tECSNe is 1.4~\% of FeCCSNe for $Z_\mathrm{CCSN}=0.004$. This increases to 1.6~\% for
$Z_\mathrm{CCSN}=0$. The constraints from \Ca{48} and \Cr{54} are similarly restrictive but to
a lesser extent, giving allowed tECSN rates between 1.6~\% and 2.7~\% of the FeCCSN rate.

In general, considering only the Zn isotopes that are produced, the solar ratios
$^A$Zn/$^{16}$O allow for larger tECSN rates. The maximum rate for which we can get an upper
limit is 52~\%, with the constraint coming from \Zn{67} for $Z_\mathrm{CCSN}=0$ and 0.001. This
reduces to 40~\% at $Z_\mathrm{CCSN}=0.004$.
The upper limits from the Zn isotopes for the zero-metallicity yields are probably so high
because of the suppression of the weak s-process in massive stars at low metallicity, where
there are less (or no) seed nuclei such as $^{56}$Fe. Indeed, using the FeCCSN yields at
$Z=0.004$, the tightest constraint from the Zn isotopes is 2~\%, coming from \Zn{66}. This is
not surprising because of how strongly \Zn{66} is produced relative to the solar abundance,
compared with the other Zn isotopes (see Figure~\ref{fig:G14a_n100_nkk}).

So, the current yields we have obtained for tECSNe suggest that tECSNe can occur at a rate of
up to $\sim1-3~\%$ of the FeCCSN rate. This is at a similar level to or approximately 1~dex
below the predictions from stellar evolution simulations convolved with a single-star IMF by
\citet{Poelarends2007,Poelarends2008}, who found that ECSNe could constitute between 3 and
21~\% of all core-collapse events \citep[see Table~3 of][]{Doherty2015}.  \citet{Doherty2015}
find lower ECSN rate predictions, so much so that the mass range for ECSNe is limited to an
initial mass interval of just $0.2~\msun$ in their simulations and results in an ECSN rate of
$2-5$~\% of all core-collapse events. This is actually in surprisingly good agreement with our
predictions using the 3d hydrodynamic simulations by \citet{Jones2016a} and computing the
nucleosynthesis from their tracer particles in a post-processing nuclear reaction network.
These stellar evolution predictions are, however, for single stars only. Those predictions
should also be taken with a pinch of salt owing to the outstanding uncertainties in the stellar
models (see the discussion in the Introduction of this paper). As we can see from
Section~\ref{sec:popsynth}, our predictions are actually also in relatively good agreement with
the rates from binary population synthesis simulations.

Interestingly, the ejected mass of $^{86}$Kr in the nucleosynthesis yields for cECSNe by
\citet{Wanajo2011} suggest that cECSNe could constitute up to 4\% of all core collapse events,
which is also in good agreement with the predictions from stellar evolution and population
synthesis. Later, \citet{Wanajo2013b} also showed that cECSNe could also be a predominant
source of \Ca{48} in addition to the rare and hypothetical class of high density SNe Ia
proposed by \citet{Woosley1997a}. Our models merge these two scenarios, where ECSNe are
the high-density SNe Ia.

\section{Isotopic ratios in pre-solar meteoritic oxide grains}

\begin{figure*}
	\centering
	\includegraphics[width=\textwidth,clip=true,trim=2cm 3.5cm 3cm 3cm]{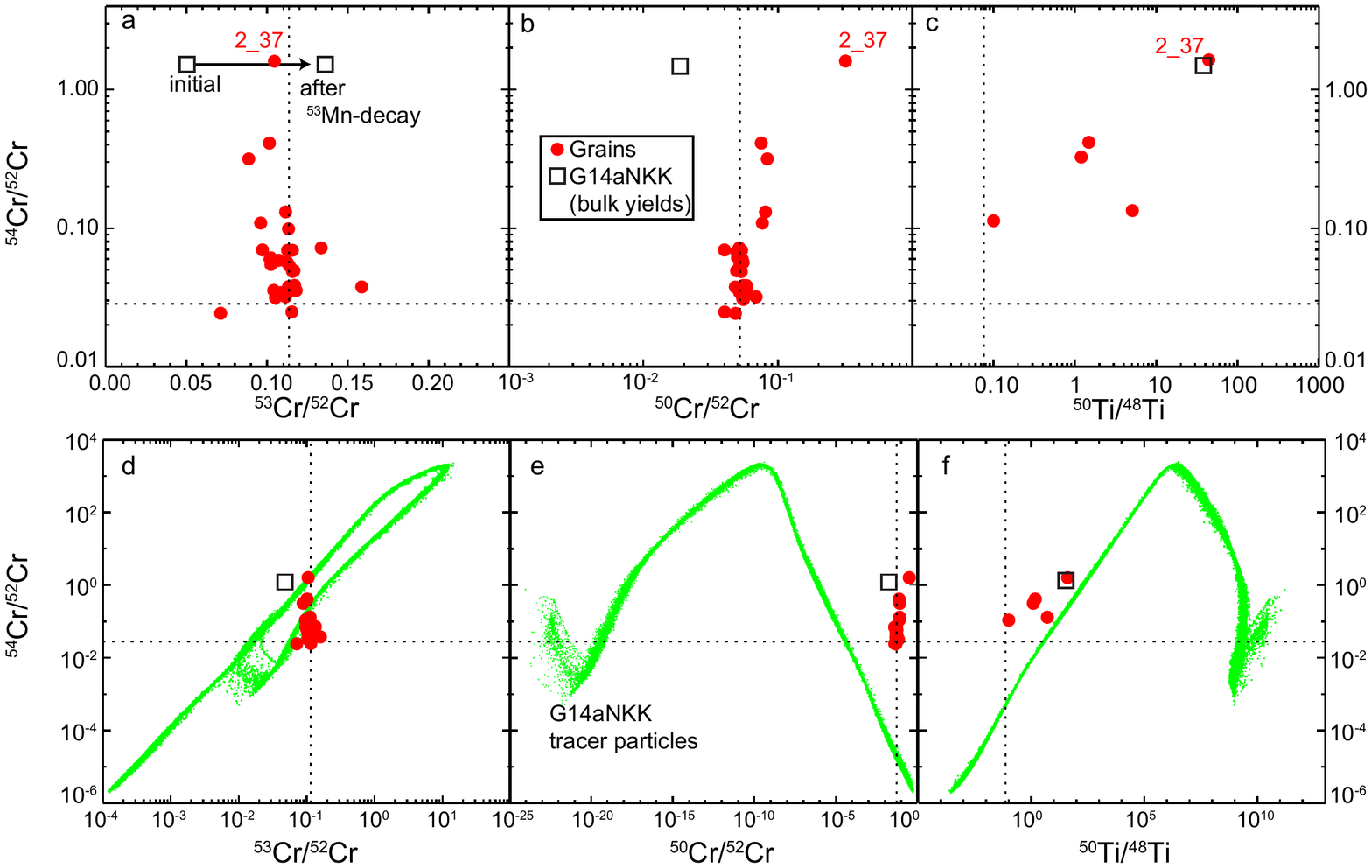}
	\caption{
	  Comparison of isotopic ratios measured in pre-solar oxide grains
	  \citep{Dauphas2010,Qin2011,Nittler2018a} compared to predictions of G14a simulation
	  with NNK weak reaction rates. Open squares are bulk yields of G14aNKK, while green
	  points (lower panels) are individual tracer particles. We note that
	  the scales are different in the upper and lower panels. The dashed
	  lines indicate solar ratios. Grain
	  \Ti{50}/\Ti{48} ratios are calculated on the assumption that all measured signal at
	  mass 50 is due to \Ti{50} \citep[i.e., \Cr{50}/\Cr{52}=0; ][]{Nittler2018a}.
	}
	\label{fig:grains}
\end{figure*}

Primitive chondritic meteorites preserve a record of the starting materials and
earliest conditions of the formation of the solar system. Among their
constituents are ``pre-solar grains,’’ nm- to $\mu$-m sized mineral grains with
extremely unusual isotopic compositions indicating that they originated in winds
and explosions of ancient stars \citep[see, e.g.][]{Hoppe2000a} and were part
of the protosolar molecular cloud. The vast majority of pre-solar grains,
including many types of silicates, oxides, carbides, and graphite, are inferred
to have formed in low-mass AGB stars or FeCCSNe. High-density SNe~Ia such as
those modelled by \citet{Woosley1997a} were suggested as the progenitors of a
small number of $\lesssim$100-nm-diameter Cr-rich oxide grains from the Orgueil
meteorite with large excesses in \Cr{54} relative to solar system materials
\citep{Dauphas2010,Qin2011}.  However, the grains in these studies were not
fully spatially resolved on sample mounts and a FeCCSN origin could not be ruled
out. Recently, \citet{Nittler2018a} reported data for several additional such
grains from Orgueil, acquired with substantially better spatial resolution.
These measurements revealed a much broader range of \Cr{54}/\Cr{52} ratios than
in previous studies, as well as resolved anomalies in \Cr{53} and/or at mass 50
in some grains. \citet{Nittler2018a} showed that large excesses at mass 50 are
most likely due to excess \Ti{50}, which could not be resolved from \Cr{50} in
these measurements. \citet{Nittler2018a} further showed that the grains’
compositions were in reasonably good agreement with the predictions of
\citet{Woosley1997a} for high-density SNe~Ia and of \citet{Wanajo2013b} for
cECSNe. It is thus useful to compare our tECSNe nucleosynthesis calculations
with the measured pre-solar grain isotopic compositions.

The Cr- and Ti-isotopic data for the \Cr{54}-rich pre-solar grains are compared
with the bulk yields of the G14a simulation with the NKK04 weak reaction rates
in Figures~\ref{fig:grains}a-c. The simulation provides an almost-perfect match
to the \Cr{54}/\Cr{52} ratio of the most extreme grain, 2-37. As seen before for
the yields of high-density SNe~Ia and cECSNe \citep{Nittler2018a}, the predicted
\Cr{50}/\Cr{52} ratio lies far below the grain data (Figure~\ref{fig:grains}b),
especially the five grains with \Cr{50}/\Cr{52}$>$0.1, all of which also have
apparent \Cr{50} enrichments (Figure~\ref{fig:grains}b). Most likely, much of
the measured signal at mass 50 in the grains is probably due to \Ti{50} rather
than to \Cr{50}. The inferred \Ti{50}/\Ti{48} ratios for these five grains,
calculated on the assumption that all measured mass-50 signal is indeed \Ti{50},
are shown in Figure~\ref{fig:grains}c.  Again, the predicted G14a bulk ejecta is
in remarkable agreement with grain 2-37 (Figure~\ref{fig:grains}c). The grains
with more modest \Cr{54} enrichments have close-to-solar \Cr{50}/\Cr{52}
ratios. Most likely the measured mass-50 signals for these grains are primarily
due to \Cr{50}, since if they were instead due to \Ti{50}, the proximity of the
data to the solar  \Cr{50}/\Cr{52} ratio would require a highly improbable
coincidence of Ti contents and \Ti{50}/\Ti{48} ratios. That said, the Cr
isotopic data for these grains are far from the G14a predictions. This  may
reflect mixing of the supernova ejecta with more solar-like material, e.g.,
circumstellar material ejected prior to the explosion. Two predicted
\Cr{53}/\Cr{52} ratios are shown in Figure~\ref{fig:grains}a, one corresponding
to directly after the explosion and one to after 0.3 Gyr, by which time all
\Mn{53} ($t_{1/2}$=3.7 Myr) has fully decayed; the composition of 2-37 lies in
between. If this grain formed in a tECSN as simulated by G14a bulk yields, this
would thus require that some of the measured \Cr{53} was originally synthesized
as \Mn{53}. To preserve the highly anomalous isotopic signatures seen without
dilution by circumstellar or interstellar matter, grain 2-37 most likely formed
within a few years of the explosion, far shorter than the lifetime of \Mn{53}.
Therefore, if a significant fraction of the observed \Cr{53} was indeed due to
\Mn{53} decay, Mn must have condensed into the grains at the time that they
formed; the G14a yields would require that grain 2-37 had a few \% stable
\Mn{55}.  Indeed, spinel minerals \citep[a likely form of the pre-solar
\Cr{54}-rich grains][]{Dauphas2010} can accommodate Mn in their structure and
future measurements of Mn in \Cr{54}-rich grains could test this hypothesis.
Alternatively, the discrepancy in \Cr{53} between the model and the data may
indicate that the ejecta was not fully mixed before the grain condensed, as
discussed further below.

\begin{figure*}
	\centering
	\includegraphics[width=\textwidth]{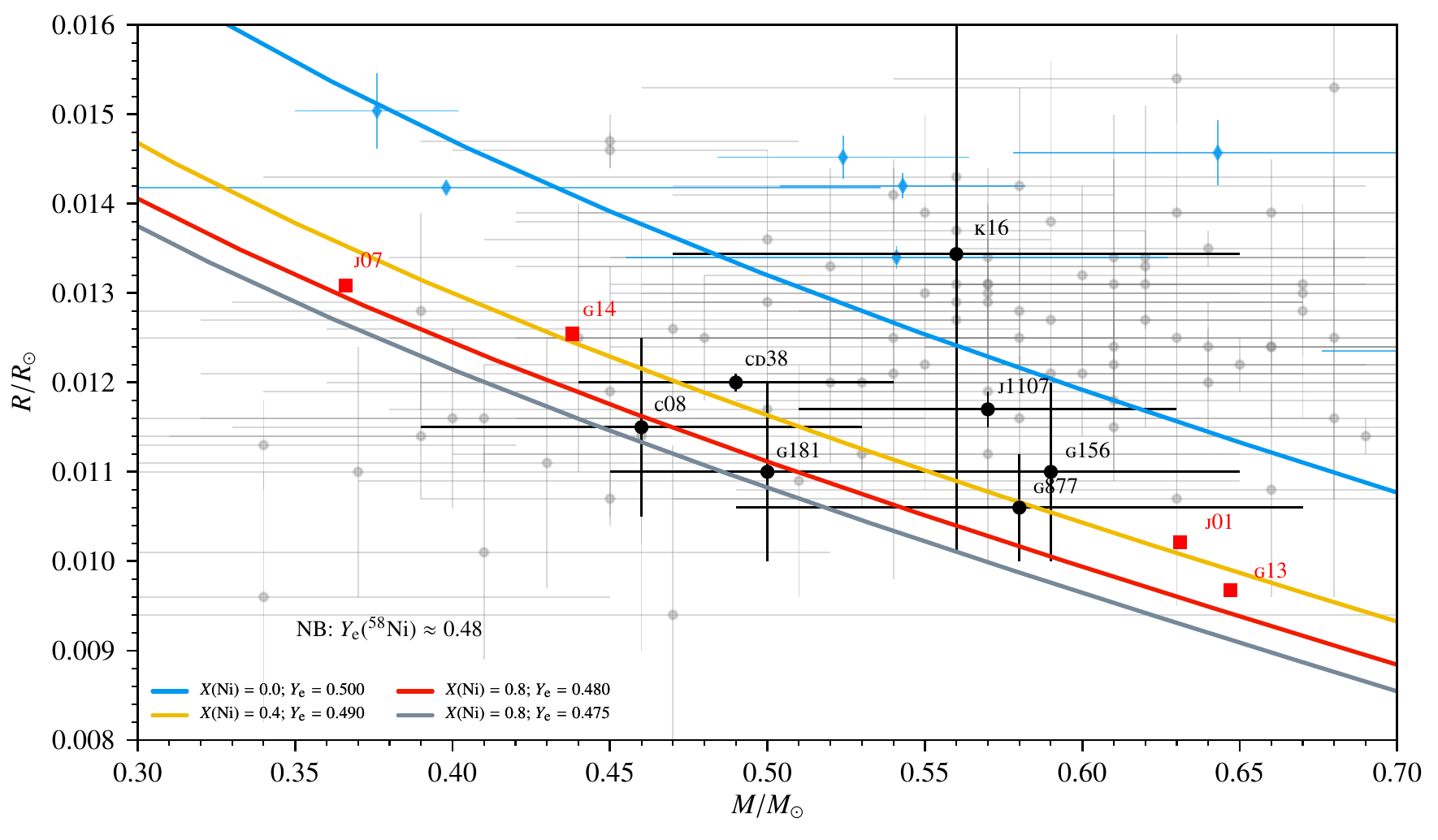}
	\caption{
		Theoretical white dwarf mass-radius relations for the surviving ONeFe bound
		remnants of the high-density ONe deflagration simulations by
		\citet{Jones2016a}. The four curves are for different ONeFe white dwarfs with
		Fe-group mass fractions $X(\mathrm{Ni})$ represented by $^{56}$Ni and $^{64}$Ni
		mixed in a ratio to give the corresponding average \ye~for the
		white dwarf. The blue curve is for pure ONe white dwarfs. The
		grey points are the measurements from \citet{Bedard2017a}, the blue points are
		measurements from \citet{Joyce2018a} which have vastly improved distance
		estimates from GAIA, and the black points
		are individual objects that have proposed as being either Fe white dwarfs or
		having Fe cores: \citet[][CD38-10980; G181-B58; G156-64]{Provencal1998a},
		\citet[][C08; WD0433+270]{Catalan2008a},
		\citet[][K16; SDSSJ124043.01]{Kepler2016a},
		\citet[][J1107; SCR J1107-342]{Bedard2017a}.
		The red points are the bound remnants from the simulations by
		\citet{Jones2016a} and this work (see Table~\ref{table:wdmr-sims}).
	}
	\label{fig:wdmr}
\end{figure*}

It is possible that the ejecta of a tECSN would not be fully mixed prior to
condensation of dust grains. To explore the range of compositions that might be
expected, the grain data are again compared to the G14a simulation in
Figures~\ref{fig:grains}d-f, only in this case the predicted compositions of
$\sim$32,000 tracer particles are shown in addition to the bulk yields. These
tracer particles contain essentially all of the ejected \Cr{54}, and about 80\%
of the total ejected Cr. The remaining tracers contain either extremely small
amounts of Cr or very \Cr{54}-poor Cr (with variable \Cr{50}) and are excluded
from the plots for clarity. Figure~\ref{fig:grains}d shows that the full range
of \Cr{53}/\Cr{52} and \Cr{54}/\Cr{52} ratios observed in the grains could be
explained by the model, if the ejecta were not fully mixed, obviating the need
to incorporate radioactive \Mn{53} in the grains. The inability of this model to
explain the grains’ measured \Cr{50}/\Cr{52} ratios is even clearer for the
tracer particles than the bulk yields (Figure~\ref{fig:grains}e). Again, the
most extreme measured mass-50 excesses most likely indicate the presence of
\Ti{50} enrichments (Figure~\ref{fig:grains}f). In this case, the tracer
particles are largely more \Ti{50}-rich than the grain compositions, perhaps
indicating a small amount of mixing with solar-like material.

In summary, the predicted Cr and Ti-isotopic compositions of the ejecta of a tECSN, as represented by 
the G14a simulation, are in remarkably good agreement with the most extreme reported \Cr{54}-rich 
pre-solar grain and the grain data as a whole can be reasonably explained by the model when individual 
ejecta tracer particles are considered. As discussed by \citet{Nittler2018a}, an ECSN origin for the 
grains is attractive in that the lifetime of the parent star (of the order of 20 Myr) is comparable to 
the timescale of star-forming regions and it may be thus more reasonable to expect an association of 
dust from such an explosion with the forming solar system than from a SN~Ia. An additional advantage of 
the present model is that, unlike the case of a cECSN \citep[e. g., ][]{Wanajo2013b}, a substantial 
amount of O is ejected by tECSN, making it more plausible for oxide grains to condense. The predicted 
O is essentially pure $^{16}{\mathrm{O}}$  and thus O-isotopic measurements of future \Cr{54}-rich 
grains may provide additional constraints on their origin.

\section{Bound ONeFe white dwarf remnants}

In the simulations that do not collapse into neutron stars, only part of the ONe core becomes
gravitationally unbound owing to energy release in thermonuclear burning, leaving behind a
gravitationally bound remnant consisting of $^{16}$O, $^{20}$Ne and some of the ashes of the
deflagration \citep{Nomoto1991,Isern1991,Canal1992,Jones2016a}. If such events do actually
occur, and occur frequently enough, then they should be represented in the Galactic white dwarf
population.

\subsection{WD mass-radius relations for bound ONeFe remnants}

We have constructed theoretical mass-radius relations for the bound ONeFe WD remnants left
behind by tECSNe. There are now several known candidate WDs that we compare with our
mass-radius relations and demonstrate could potentially be the gravitationally bound remnants
of these explosions.

Using the equation of state by
\citet{PotekhinPC2010}\footnote{\href{http://www.ioffe.ru/astro/EIP/}{http://www.ioffe.ru/astro/EIP/}}
we have constructed hundreds of spherically symmetric isothermal, uniform-composition white
dwarf models in hydrostatic equilibrium using a Cash-Karp type Runge-Kutta integrator
\citep{Cash1990a} starting from a given central density and integrating outwards to the
surface, from which we have constructed theoretical white dwarf mass--radius relations for the
bound remnants. The WDs are assumed to have no H or He layer at the surface. A range of
compositions are possible outcomes from the hydrodynamic simulations, characterized by some
fraction of Fe-group isotopes and an average \ye~\citep[see][their Table~1]{Jones2016a}. We
have therefore chosen to use a two-parameter model for the white dwarf composition, where the
ratio $X(^{16}\mathrm{O})/X(^{20}\mathrm{Ne})$ is held constant at $0.65/0.35=1.86$ (i.e. the
same as the initial conditions before the deflagration) and the mass fraction of Ni and the
\ye~are varied. We have assumed for simplicity that the Ni is made up from the two isotopes $^{56}$Ni
and $^{64}$Ni, whose ratio is determined by \ye. More explicitly, given
an ``Fe-group'' mass fraction $X_\mathrm{Ni}$ and an average \ye, the composition is given by
\begin{eqnarray}
	X_{56} = 8(2\ye-1) + X_\mathrm{Ni}, \\
	X_{64} = X_\mathrm{Ni} - X_{56} = -8(2\ye-1), \\
	X_{16} = 0.65(1-X_\mathrm{Ni}), \\
	X_{20} = 1 - X_\mathrm{Ni} - X_{16}.
\end{eqnarray}

We plot the resulting mass--radius curves for $(X_\mathrm{Ni},\ye) =
\{(0,0.5),(0.4,0.49),(0.8,0.48),(0.8,0.475)\}$ in Figure~\ref{fig:wdmr}.  The
blue curve is for pure ONe white dwarfs. We have also plotted in
Figure~\ref{fig:wdmr} the measurements from \citet{Bedard2017a} (grey points)
and some individual objects that have been proposed to be either Fe white dwarf
or Fe-core white dwarf candidates: \citet[][CD38-10980; G181-B58;
G156-64]{Provencal1998a}, \citet[][C08; WD0433+270]{Catalan2008a}, \citet[][K16;
SDSSJ124043.01]{Kepler2016a}, \citet[][J1107; SCR J1107-342]{Bedard2017a}.
Several of the individually-named (black points) candidates are reasonably fit
with the cold ONeFe WD mass-radius relations. Only K16 appears to be more
consistent with an ONe WD, although its error bars are quite large and all of
our theoretical mass-radius curves the ONeFe WDs pass through the error bars for
K16. The cloud of grey points from \citet{Bedard2017a} contain several
candidates that could be cold ONeFe WDs according to our theoretical mass-radius
relations; some of the extreme WDs (in the lower-left portion of the figure) do
not appear to be consistent with an ONeFe WD although again the error bars are
quite large. We note at this point that observational tests of the WD
mass-radius relationship are subject to uncertainties in the distance and
surface gravity measurements. Previously, the distance estimates provided the
largest source of the uncertainty, but with the launch of the GAIA mission the
distance uncertainties have been considerably reduced and the spectroscopic
measurements of H lines (from which the surface gravity can be derived) now pose
the largest uncertainty \citep[see, e.g.][]{Joyce2018a}. Data from
\citet{Joyce2018a} using GAIA parallax distances are included as blue points in
Figure~\ref{fig:wdmr} -- note the substantially reduced radius error bars.  All
of the WDs reported by \citeauthor{Joyce2018a} appear to be more consistent with
ONe or CO WDs (which would lie in the upper right of the figure) than ONeFe WDs.

Also plotted in Figure~\ref{fig:wdmr} are some of the models from \citet[][red
squares]{Jones2016a}. The relevant properties (i.e. remnant masses, mass
of Fe-group elements and average electron fraction) from Table~1 of
\citet{Jones2016a} are repeated in Table~\ref{table:wdmr-sims} for convenience.
The bound ONeFe WD remnants do not match particularly well with any individual
observed candidate, although they do populate a similar portion of the
mass-radius plane. The simulations G13, G14, J01 and J07 shown in
Figure~\ref{fig:wdmr} did not include Coulomb corrections in the EoS. Models
including these corrections yielded significantly larger bound remnant masses
(see Table~\ref{table:wdmr-sims}) and would be outside the domain of this figure
to the lower right. There are also some white dwarf candidates with such larger
masses reported by \citet{Vennes2017a} that may be good fits for those
simulations.

\begin{table}
	\caption{Decayed mass fractions of elements in the bound ONeFe remnant and the ejected material
		of the G14 simulation from \citet{Jones2016a}. Decays were performed over
		$10^{16}$~s.}
	\label{table:G14elements}
	\centering
	\begin{tabular}{l c c c}
		\toprule\toprule
		Element  & Z  & ejecta        & bound remnant \\
		\midrule
		H        & 1  & 1.05e-09      & 2.30e-10 \\
		He       & 2  & 6.25e-05      & 3.58e-05 \\
		Li       & 3  & 3.81e-16      & 1.41e-15 \\
		B        & 5  & 3.98e-15      & 3.78e-15 \\
		C        & 6  & 4.74e-08      & 4.91e-08 \\
		N        & 7  & 9.97e-08      & 1.29e-07 \\
		O        & 8  & 3.68e-01      & 4.58e-01 \\
		F        & 9  & 4.11e-12      & 5.78e-12 \\
		Ne       & 10 & 1.85e-01      & 2.29e-01 \\
		Na       & 11 & 2.68e-08      & 3.64e-08 \\
		Mg       & 12 & 2.98e-03      & 3.91e-03 \\
		Al       & 13 & 2.30e-06      & 3.12e-06 \\
		Si       & 14 & 2.34e-02      & 2.87e-02 \\
		P        & 15 & 3.13e-06      & 4.18e-06 \\
		S        & 16 & 1.45e-02      & 1.75e-02 \\
		Cl       & 17 & 3.17e-06      & 3.61e-06 \\
		Ar       & 18 & 3.31e-03      & 3.93e-03 \\
		K        & 19 & 8.74e-07      & 1.08e-06 \\
		Ca       & 20 & 5.38e-03      & 6.54e-03 \\
		Sc       & 21 & 2.08e-08      & 2.47e-08 \\
		Ti       & 22 & 4.20e-03      & 1.94e-03 \\
		V        & 23 & 2.09e-04      & 1.03e-04 \\
		Cr       & 24 & 1.57e-02      & 7.92e-03 \\
		Mn       & 25 & 7.09e-03      & 3.83e-03 \\
		Fe       & 26 & 3.08e-01      & 1.94e-01 \\
		Co       & 27 & 5.19e-04      & 2.65e-04 \\
		Ni       & 28 & 5.35e-02      & 3.03e-02 \\
		Cu       & 29 & 9.99e-05      & 6.05e-05 \\
		Zn       & 30 & 6.80e-03      & 6.82e-03 \\
		Ga       & 31 & 2.05e-05      & 2.64e-05 \\
		Ge       & 32 & 1.29e-04      & 7.89e-04 \\
		As       & 33 & 4.37e-06      & 7.89e-06 \\
		Se       & 34 & 4.80e-04      & 4.12e-03 \\
		Br       & 35 & 3.13e-05      & 1.85e-04 \\
		Kr       & 36 & 3.95e-04      & 2.28e-03 \\
		Rb       & 37 & 1.30e-05      & 2.48e-05 \\
		Sr       & 38 & 5.84e-08      & 3.92e-07 \\
		Y        & 39 & 2.39e-09      & 6.56e-09 \\
		Zr       & 40 & 8.60e-10      & 1.62e-09 \\
		Nb       & 41 & 9.70e-14      & 5.28e-13 \\
		Mo       & 42 & 8.43e-15      & 7.92e-14 \\
		Tc       & 43 & 1.26e-42      & 8.20e-43 \\
		Ru       & 44 & 4.29e-17      & 1.80e-17 \\
		\bottomrule \bottomrule
	\end{tabular}
\end{table}

There are other WD candidates identified by \citet{gaensicke2010a} and \citet{raddi2018a} worth
mentioning here. Unfortunately, the data for the WDs identified by \citet{gaensicke2010a} are
insufficient to derive the WD mass, however their large O/C ratios imply that they are, or
were, ONe WDs as opposed to CO WDs.

The WD LP~40-365 studied by \citet{raddi2018a} is estimated to have a radius of
0.18$\pm0.01~R_\odot$, and a mass of 0.37$^{+0.29}_{-0.17}$~\msun, placing it outside of the
domain of Figure~\ref{fig:wdmr}.  This means that the white dwarf has more than ten times the
radius that we would expect it to have from our hydrostatic cold ($2\times10^4$~K) WD models of
the bound ONeFe remnant. For the radius of the WD to be this large, the star would need to be
substantially hotter, say $\sim 10^7$~K.

\subsection{Atmospheric composition of LP~40-365}

\citet{raddi2018a} were able to spectroscopically derive compositional information for several
elements in the atmosphere of LP~40-365, and they compared the composition to published yields
from CCSN simulations, SN~Ia simulations (DDTs) and SN~Iax simulations (pure deflagrations
with/without hybrid C/O/Ne progentiors). The detection of Mn in the atmosphere suggests that
the composition originated in a single-degenerate SN~Ia \citep{seitenzahl2013b}, and the fact
that a WD still exists suggests that the explosion failed to gravitationally unbind the entire
star \citep{kromer2013a}.

\begin{figure}
	\centering
	\includegraphics[width=0.5\textwidth]{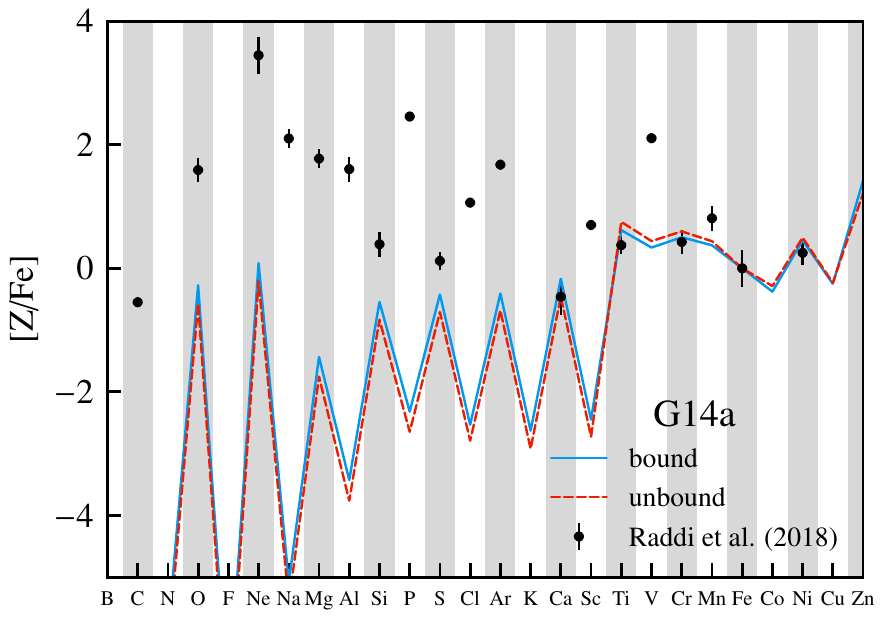}
	\caption{
		Elemental composition ratios ratio of the bound remnant and the unbound ejecta
		of the G14 simulation with respect to Fe and relative to the solar ratio. The
		black points are the spectroscopically determined composition of the atmosphere
		of the white dwarf LP 40-365 from \citet{raddi2018a}. While the Ca, Ti, Cr, Fe and
		Ni abundances relative to Fe fit very well, the lighter elements are in
                distinct tension with the measurements.
	}
	\label{fig:raddiWD}
\end{figure}

Both of these characteristics ([Mn/Fe]$>0$ and a gravitationally bound remnant) are shared by
our ONe deflagration simulations of tECSNe. In Table~\ref{table:G14elements} we give the
decayed mass fractions of elements in the bound remnant and in the ejecta of the G14
simulation. We plot the ratios of the decayed elemental composition to Fe compared to the solar
ratios for the bound remnant and the ejecta of simulation G14 in Figure~\ref{fig:raddiWD}. The
ratios for Ca, Ti, Cr, Fe and Ni in the G14 simulation appear to fit the data very well. The
simulation produces a super-solar ratio for Mn of [Mn/Fe]$=0.44$, which is approximately half
that measured in LP~40-365 ([Mn/Fe]$=0.82\pm0.18$) by \citeauthor{raddi2018a} and outside of
the error bars, but not wildly inconsistent. Sc and V present much larger tensions with the
measurements of higher-mass elements and stand out as being the only obviously problematic
elements heavier than Ca. For the rest of the intermediate-mass elements the agreement between
G14 and LP~40-365 is very poor, and even for Ne the simulation is 3~dex below the observational
data. One of the caveats of our current nucleosynthesis simulations is the assumption that the
initial composition is a mixture of pure $^{16}$O and $^{20}$Ne. This means that there is no Na
or Mg, etc from the prior C burning phases. There is also no signature of the metals that would
have been present in the cloud that the star formed from. Lastly, ECSNe should be most
prevalent from stars in binary systems (see Section~\ref{sec:popsynth}) in which we expect a He
shell and/or H envelope to surround the ONe core, which could further influence the atmospheric
composition of the bound ONeFe remnant from a tECSN. Accounting for these shortcomings could
help to alleviate some of the tensions that our tECSN simulations have with the atmospheric
composition of LP~40-365, however it seems unlikely that the large discrepancies in the light-
and intermediate-mass elements can be completely resolved in this way.

\section{Discussion of implications and concluding remarks}

We have studied the nucleosynthesis in the ejecta and the bound ONeFe remnants of the
thermonuclear ECSNe (tECSNe) from \citet{Jones2016a}. The ejecta contains very large abundances
of the neutron-rich isotopes \Ca{48}, \Ti{50}, \Cr{54} and \Zn{66} relative to solar. When weak
reaction rates for \emph{fp}- and \emph{fpg}-shell nuclei (to the neutron-rich side of the
\emph{pf} shell) are included, the abundances of \Ca{48} and \Zn{66} are enhanced in the
ejecta, the abundances of \Ti{50} and \Cr{54} are reduced and isotopes of the trans-iron
elements Ge, Se and Kr are produced in greater abundance. The yields share many similarities
with the core-collapse ECSN (cECSN) simulations by \citet{Wanajo2013b} and the high-density SNe
Ia simulations by \citet{Woosley1997a}. In the tECSNe scenario we present, ECSNe are the
high-density SNe Ia\footnote{Technically, the SN class will of course depend on the light curve
	and spectrum of tECSNe, which will depend greatly on how much of the envelope, if any,
	remains when the star explodes.}.

The ejecta exhibits a high [Zn/Fe] ratio, which makes it an interesting candidate for
explaining the high [Zn/Fe] in the early Milky Way, for which hypernovae are currently the most
favourable scenario. The high [Ti/Fe] and [Mn/Fe] in the ejecta, if injected at early times
into the Milky Way, could also help to alleviate the current tensions of GCE models with the
observations of these two elements.

Owing to the low electron fractions achieved in portions of the ejecta, the
$2.81\times10^{-3}~\msun$ yield of $^{60}$Fe is quite large -- approximately ten or more times
that of a FeCCSN. Perhaps more interestingly, owing to the different origin of $^{60}$Fe in the
tECSN scenario as compared with a FeCCSN, the molar $^{60}$Fe/$^{26}$Al abundance ratio in the
ejecta $Y(^{60}\mathrm{Fe})/Y(^{26}\mathrm{Al}) = 4.94\times10^4$, which is $4-5$ orders of
magnitude larger than what is expected from massive stars and FeCCSNe, which has interesting
implications for interpreting the line ratio of 0.17 measured in the diffuse ISM by
INTEGRAL/SPI. If the $^{26}$Al from the progenitor envelope is included in the yield, the ratio
in the tECSN is $Y(^{60}\mathrm{Fe})/Y(^{26}\mathrm{Al}) \approx 130$, which is lower but still
four orders of magnitude greater than the ISM value.

Using the solar abundance distribution and the FeCCSN yields from \citet{Nomoto2006}, we place
an upper limit on the occurrence of tECSNe to approximately $1-3~\%$ of the FeCCSN rate. This
is in good agreement with the predictions from stellar evolution modelling and population
synthesis simulations, which give $2-20~\%$ and $3-4~\%$, respectively. This is a somewhat
surprising result and means that potentially all ECSNe being thermonuclear explosions does not
apparently introduce an inconsistency between stellar evolution, binary population synthesis
and galactic chemical evolution.  If all ECSNe/AIC were tECSNe, this would mean that the Crab
nebula is not the remnant of an ECSN \citep{Davidson1982,Nomoto1982crabnature,Smith2013a}.
Indeed, \citet{Woosley2015a} have shown that low-mass FeCCSNe could also be valid formation
scenario for the Crab nebula and pulsar and \citet{Gessner2018a} demonstrate that the kick
velocity of the Crab pulsar is more consistent with a low-mass FeCCSN than a cECSN.  The
outcome of the ONe deflagration is so sensitive to the prior evolution leading up to the
$^{20}$Ne electron capture phase and to the nuclear reaction rates themselves, amongst other
things, that it is not impossible that both collapses and partial thermonuclear explosions
could occur.

If accreting ONe WDs in ultra-close binaries undergoing stable mass transfer and retaining mass
eventually undergo AIC but do not collapse into neutron stars (but are instead thermonuclear
explosions), then they would no longer be candidates for forming low mass black holes
\citep[BHs, e.g.][]{belczynski2004a}. It is indeed currently the case that these low mass BHs
have not been observed in binary systems, which is consistent with the scenario that AIC events
do not produce NSs.

The isotopic ratios \Cr{54}/\Cr{52} and \Ti{50}/\Ti{48} (and \Cr{53}/\Cr{52} if
the grains condense before mixing with the ISM) in a sub-set of meteoritic
pre-solar oxide grains that have been identified as having extremely
large \Cr{54} and \Ti{50} abundances are able to be very well reproduced by our
tECSN simulations. The agreement is quite remarkable in fact. The
close-to-solar \Cr{50}/\Cr{52} ratios measured in less
anomalous (though still \Cr{54}-enriched) grains, on the other hand, are much
more difficult to match with the tECSN simulations and may require
mixing of the ejecta with unprocessed pre-supernova material. tECSNe are very
good candidates for explaining these types of oxide grains because there is a
substantial amount of O in the ejecta, which is not the case for the yields of
cECSNe.

The bound ONeFe WD remnants that tECSNe are expected to leave behind also look to be consistent
with several observed candidate WDs. Theoretical mass-radius relation curves computed with
typical remnant compositions pass through the error bars for several such objects.
Unfortunately, for one particular object LP~40-365 where there is a
spectroscopically-determined elemental composition for the WD's atmosphere our WD
remnants are far too small. This could be remedied if the remnants were hotter (about
$10^7$~K), but even then we are unable to explain the entire composition in a satisfactory
manner. For a sub-set of the elements though, including Ca, Ti, Cr, Mn, Fe and Ni, our model
does match very well.  Much more accurate parallax distances of WDs are available with the GAIA
mission, making the spectroscopic determination of the surface gravities of WDs now the most
uncertain aspect of constraining the observed WD mass-radius relation, which should help in
either confirming or denying whether some or all ECSNe are tECSNe.

The rate predictions made in this paper using 3D hydrodynamics simulations and nucleosynthesis
are fortunately not plagued by the difficult challenges of modelling the TP-SAGB phase of
super-AGB stars or the convectively-bounded flames of low-mass FeCCSN progenitors, however they
do have their own, sizeable, baggage attached. This includes the accuracy of the \emph{fp}- and
\emph{fpg}-shell nuclear data, initial conditions for the modelling of the deflagration front
and the precise ignition density of the deflagration. Many of these uncertainties are
adequately discussed by \citet{Jones2016a} to which the interested reader is referred for
further reading.

\begin{acknowledgements}

	This work was supported by the US Department of Energy LDRD program
	through the Los Alamos National Laboratory. Los Alamos National
	Laboratory is operated by Triad National Security, LLC, for the National
	Nuclear Security Administration of U.S. Department of Energy (Contract
	No. 89233218NCA000001). SJ acknowledges support from a Director's
	Fellowship at Los Alamos National Laboratory and thanks Robert Fisher,
	Enrique Garcia-Berro and Ken Shen for stimulating discussion and
	pointing out several interesting white dwarf candidates, and Shinya
	Wajano for providing the complete yields from his ECSN simulations.  SJ
	and FKR acknowledge support from the
	\href{https://www.klaus-tschira-stiftung.de}{Klaus Tschira Stiftung}.
	The work of FKR was supported by the German Research Foundation (DFG)
	via the Collaborative Research Center SFB~881 ``The Milky Way System''.
	A.~J.~R.~ is supported by the Australian Research Council through grant
	number FT170100243.  I.~R.~S.~was supported by Australian Research
	Council Grant FT160100028.  R.~R.~has received funding from the European
	Research Council under the European Unions's Seventh Framework Programme
	(FP/2007-2013) / ERC Grant Agreement n.  615126.  M.~P.~acknowledges the
	support of STFC through the University of Hull Consolidated Grant
	ST/R000840/1 and from the ERC Consolidator Grant (Hungary) funding
	scheme (project RADIOSTAR, G.A. n. 724560).

\end{acknowledgements}



\end{document}